\documentclass[aoas,preprint]{imsart}

\RequirePackage[OT1]{fontenc}
\RequirePackage{amsthm,amsmath}
\RequirePackage[numbers]{natbib}
\RequirePackage[colorlinks,citecolor=blue,urlcolor=blue]{hyperref}
\usepackage{tikz}
\usetikzlibrary{arrows}

\arxiv{arXiv:0000.0000}

\startlocaldefs
\numberwithin{equation}{section}
\theoremstyle{plain}

\endlocaldefs

\begin{document}

\begin{frontmatter}

\title{Frequency behaviour for multinomial counts of fisheries discards via a nested wavelet zero and $N$ inflated binomial model}
\runtitle{Frequency behaviour for fisheries discards}

\begin{aug}
  \author{\fnms{Andrew C.}  \snm{Parnell}\corref{}\ead[label=e1]{Andrew.Parnell@ucd.ie}},
  \author{\fnms{Norman} \snm{Graham}\ead[label=e2]{norman.graham@marine.ie}},
  \author{\fnms{Andrew L.} \snm{Jackson}\ead[label=e3]{andrew.jackson@tcd.ie}},
  \and
  \author{\fnms{Mafalda}  \snm{Viana} \ead[label=e4]{mafalda.viana@glasgow.ac.uk}}

  \runauthor{A.C. Parnell et al.}

  \affiliation{University College Dublin, Marine Institute of Ireland, Trinity College Dublin, and University of Glasgow}

  \address{School of Mathematical Sciences (statistics), Complex and Adaptive Systems Laboratory,\\ University College Dublin,\\ 
          \printead{e1}}
  \address{Fisheries Sciences Services, Marine Institute, Rinville, Oranmore, Co. Galway, Ireland,\\ 
          \printead{e2}}
  \address{Department of Zoology, School of Natural Sciences, Trinity College Dublin, Dublin 2, Ireland,\\ 
          \printead{e3}}
  \address{Institute of Biodiversity, Animal Health and Comparative Medicine, University of Glasgow,\\ 
          \printead{e4}}
\end{aug}

\begin{abstract}
In this paper we identify the changing frequency behaviour of multinomial counts of fish species discarded by vessels in the Irish Sea. We use a Bayesian hierarchical model which captures dynamic frequency changes via a shrinkage model applied to wavelet basis functions. Wavelets are known for capturing data features at different temporal scales; we use a recently-proposed shrinkage prior from the factor analysis literature so that features at the finest levels of detail exhibit the greatest shrinkage. Rather than using a multinomial distribution for monitoring the changes in discards over time, which can be slow to fit and inflexible, we use a nested zero-and-N inflated (ZaNI) binomial distribution which enables much faster computation with no obvious deterioration in model flexibility. Our results show that seasonal behaviour in these data are not regular and occur at different frequencies. We also show that the nested ZaNI binomial distribution is a good fit to multinomial count data of this sort when an informative nested structure is applied. \\
\end{abstract}

\begin{keyword}[class=MSC]
\kwd{62H99}
\kwd{60E05}
\kwd{62J02}
\kwd{62P12}
\kwd{92D40}
\end{keyword}

\begin{keyword}
\kwd{Wavelets}
\kwd{Multinomial}
\kwd{Nested Binomial}
\kwd{Zero Inflation}
\kwd{Hamiltonian Monte Carlo}
\end{keyword}

\end{frontmatter}

\section{Introduction}
The discarding of fish at sea is an issue of European level prominence \cite{Commission2011}. The mortality rate of the discarded organisms is very high, often approaching 100\% \citep{Alverson1994}, so this practice brings many economic and ecological concerns \citep{Clucas1997,Kelleher2005}. These include: loss in biodiversity, increased unreported fishing mortality, and alterations of predator-prey relationships \citep{VanOpzeeland2005,Votier2004}. Such issues are of particular concern in fisheries with a high discards rate such as Nephrops fisheries which typically discard approximately 60\% of their total catch \citep{Kelleher2005}. However, very little is known about discards and its dynamics. Here we use a discards dataset to illustrate our proposed methodology. Our key interest is the changing frequency dynamics of discards over time.\\

The temporal dynamics of discards may be of interest for both retrospective and prospective monitoring. Retrospectively, because compositional change may be a consequence of some primary stressor such as fishing strategies, policy changes or market pressures (e.g. non-marketable species are always discarded) and its dynamics can be tracked \citep{Tsagarakis2008}. On the other hand, this information can be used prospectively to identify key species that should be monitored closely or assessed in combination with other species, and thus become an input into discard management plans. A further primary interest is that discards will be banned by the EU from 2015/16\footnote{\url{http://ec.europa.eu/fisheries/reform/docs/discards_en.eps}}. This study provides a means of comparing pre- and post-ban mortality.\\

The fisheries data we use are multivariate counts of different numbers of species discarded at sea.  The counts are constrained because the boats on which they are caught are limited by time, quota availability, market demands and the perishable nature of the catch. The traditional method for modelling such data is to use the multinomial distribution and to place a time series structure on the (transformed) individual proportions. This may not be optimal, for reasons we discuss in Section \ref{multinomial}. We instead use a nested binomial decomposition of the likelihood which factorises and allows us to fit multiple species simultaneously and in parallel without the need to treat each species' count as independent. Since the individual counts of species discards are likely zero-inflated, the nesting structure induces further zero-and $N$ inflation. We use a novel Zero and $N$-inflated (ZaNI) binomial distribution to counteract these issues, and place further structure so that the zero/$N$ inflation is itself a function of the mean and seasonality of the process. We include random effects to allow for repeated measures which are present in our data through the multiple fishing trips that each fishing vessel undertakes.\\

Our paper is structured as follows. In Section \ref{wavelets} we review wavelet methods and discuss the shrinkage approaches we use. In Section \ref{multinomial} we cover multinomial time series modelling and outline our nested ZaNI-binomial distribution. Section \ref{data} covers our case study data set. Section \ref{model} presents the full detail of our model, and Section \ref{results} the resulting output. We conclude in Section \ref{discussion}. All data and code are available from \url{http://mathsci.ucd.ie/~parnell_a/}. \\

\section{Wavelet smoothing and dynamic frequency identification}\label{wavelets}

Discovering changes in frequency behaviour over time is a challenging problem and covers a diverse area of literature. A  recent review \cite{So2014} shows that many methods have been used to uncover dynamic seasonality structure. These approaches can be broadly categorised as:  
\begin{itemize}
\item Holt-Winters smoothing methods \cite{Holt2004,Winters1960} which use the framework of exponential smoothing,
\item Seasonal ARIMA models \citep[SARIMA,][]{Box2008}, which follow the Box-Jenkins school of time series modelling, and 
\item Periodic Auto-Regressive models \citep[e.g.][]{PhilipHansFransesAuthor2004} which, similar to our approach,  remove the need to explicitly separate trend and seasonality components.
\end{itemize}
Other authors \citep[e.g.][]{Bildik2001,So2014} have developed methods that are both seasonal in mean and in variance. We do not explore seasonal variance models here though as our transformed proportions will naturally account for changing variance with seasonality.\\

\begin{figure}[!h]
\includegraphics[width=\textwidth]{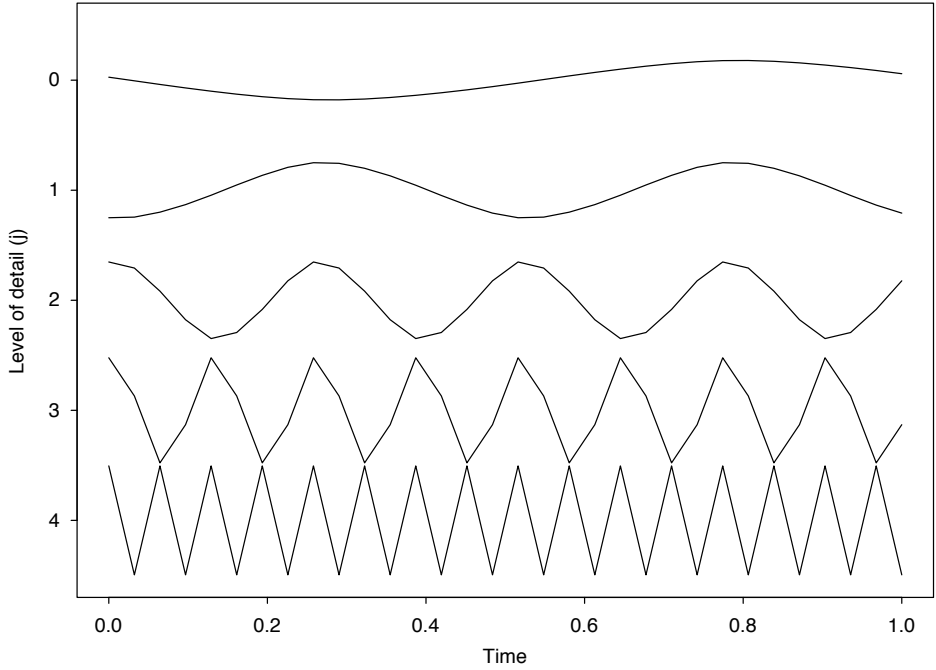}
\caption{A plot of the unweighted least asymmetric Daubechies' wavelet basis functions. The y-axis denotes the level of detail. Higher levels of details denote finer structure. In our shrinkage prior, we penalise these higher levels of detail.}
\label{UnweightedWavelets}
\end{figure}

Our approach fundamentally differs from those above in that we are using a smoothing-type approach to model time series data in an attempt to explain the changing seasonality in observational data, rather than using the model for forecasting. One such smoothing approach for identifying dynamic seasonality is that of wavelets. Wavelet methods have been widely discussed; see \cite{Nason2008}, or \cite{PeterMullerEditor1999} for a Bayesian version. In the Bayesian paradigm they are most commonly used as a prior distribution over functions, consisting of a series of orthonormal basis functions which have successively finer support. Their main advantage is that they can represent variation at different levels of detail, commonly presented in a wavelet transform (Figure \ref{UnweightedWavelets}).\\

 Standard wavelet theory \cite{Nason2008} provides that any square integrable function can be written as:
\begin{align}
\mu(t) &= \sum_k \beta_k \phi_k(t) + \sum_{j>0} \sum_k \gamma_{jk} \psi_{jk}(t)
\label{wavelet_eqn}
\end{align}
where $j,k \in \mathcal{Z}$, parameters $\theta = (\beta,\gamma)$ and wavelet scaling functions $\phi_k,\psi_{jk}$, where $\psi_{jk}(t) = 2^{j/2} \psi(2^jt-k)$ and $\phi_k(t) = \phi(t-k)$. $\psi$ and $\phi$ are known as wavelet and scaling functions respectively at level of detail $j$ and shift $k$. A key choice is that of $\psi$ and $\phi$; a number of common families have been proposed which yield suitable behaviour. The most famous is perhaps the Haar wavelets, though we instead use Daubechies' Least Asymmetric \citep[see][for full details of these terms]{Nason2008} as these most closely resemble a Fourier basis and are thus best suited for identifying seasonality. See Figure \ref{UnweightedWavelets} for a plot of the wavelet basis functions we use.\\

Once the wavelet family has been chosen we can create a square matrix $W$ of wavelet basis functions evaluated on a grid of size $2^d$ for a suitably chosen $d$ which represents the maximum level of detail. We then write $\mu = HW\theta$ where $H$ is a spline interpolation matrix to match the irregular time observations on to the $2^d$ grid, and $\theta$ are parameters representing the weights on the basis functions. To proceed, it is common to place a prior distribution on $\theta$ so that the lowest levels of detail are shrunk more than those at coarser levels. Common techniques for shrinking the values of $\theta$ include \cite{PeterMullerEditor1999} who use a version of $L_0$ shrinkage so that $\gamma_{jk}$ in Equation \ref{wavelet_eqn} above is replaced by $s_{jk} \gamma_{jk}$ where $\pi(s_{jk}=1) = \alpha^j$ where $j$ is the level of detail. Thence $\gamma_{jk}|s_{jk}=1 \sim N(0,\tau r_j)$ where $\tau$ is a variance parameter to be estimated and $r_j=2^{-j}$ thus providing a further restriction on the size of the variance at each successive level of detail. Other methods for Bayesian wavelet shrinkage can be found in \cite{Nason2008}.\\

The technique we use for wavelet shrinkage was proposed and used in the Bayesian factor analysis literature by \cite{Bhattacharya2011}. We write $\theta_l$ as the basis function for $l=1,\ldots,2^d$, associated with detail level $d(l)$. We place the prior $\theta_l \sim N\left(0,(\phi \tau_{d(l)})^{-1}\right)$ where $\phi$ is a global shrinkage parameter and $\tau_{d(l)}$ a local shrinkage parameter specific to that detail level. The detail level shrinkage is obtained by setting $\tau_{d(l)}|\delta_{d(l)} = \prod_{s=1}^{d(l)} \delta_s$ where $\delta_{1} \sim Ga(\alpha_1,1)$ and $\delta_{d(l)>1} \sim Ga(\alpha_2,1)$ with $\alpha_1>0$ and $\alpha_2>1$. The cumulative multiples of $\delta$  which all have mean $>1$ have the effect of reducing the prior variance at each successive level of detail. $\alpha_1$ and $\alpha_2$ are estimated as part of the model. We discuss prior choices for these values in Section \ref{model}.\\

\begin{figure}[!h]
\begin{tabular}{cc} \includegraphics[width=0.5\textwidth]{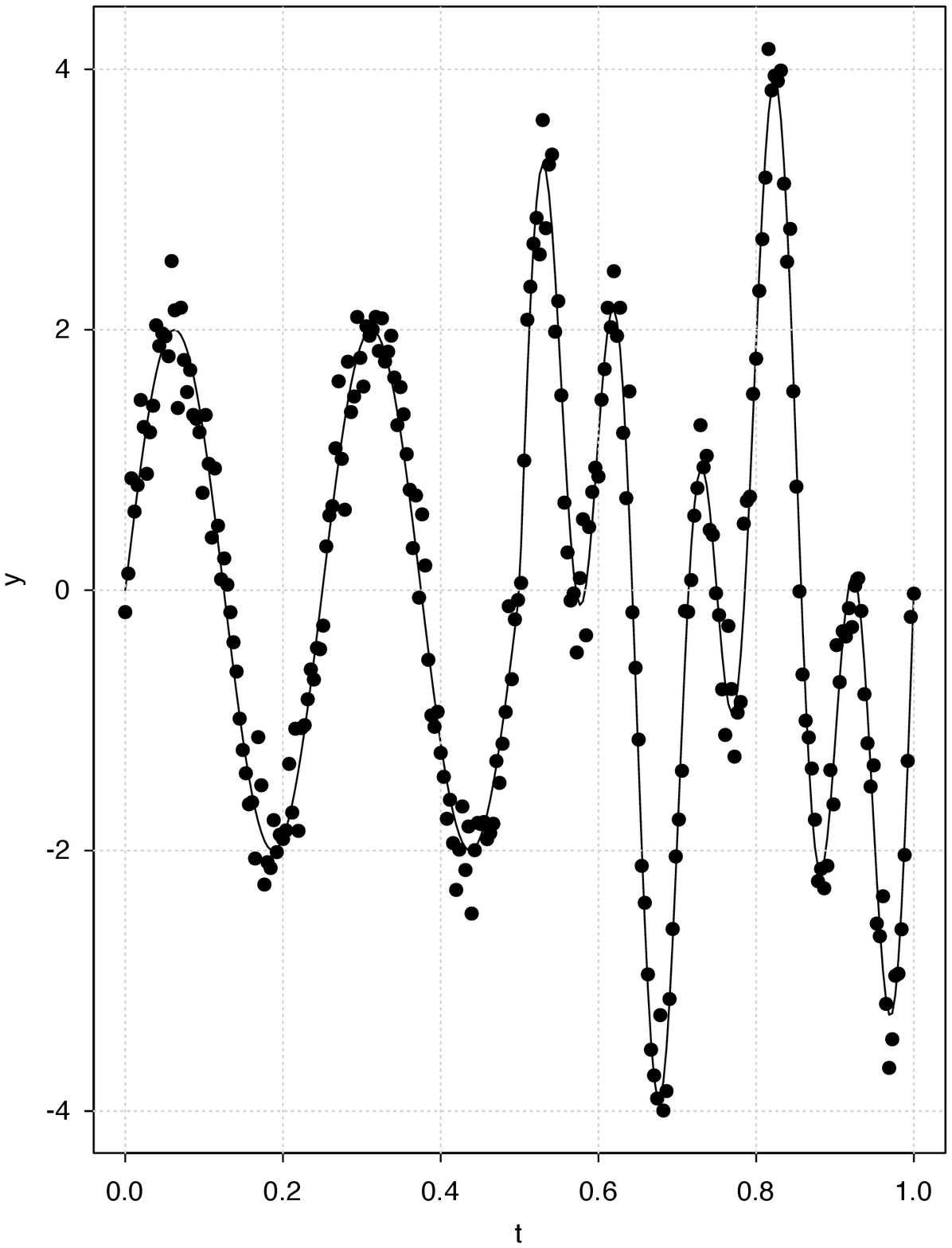} & \includegraphics[width=0.5\textwidth]{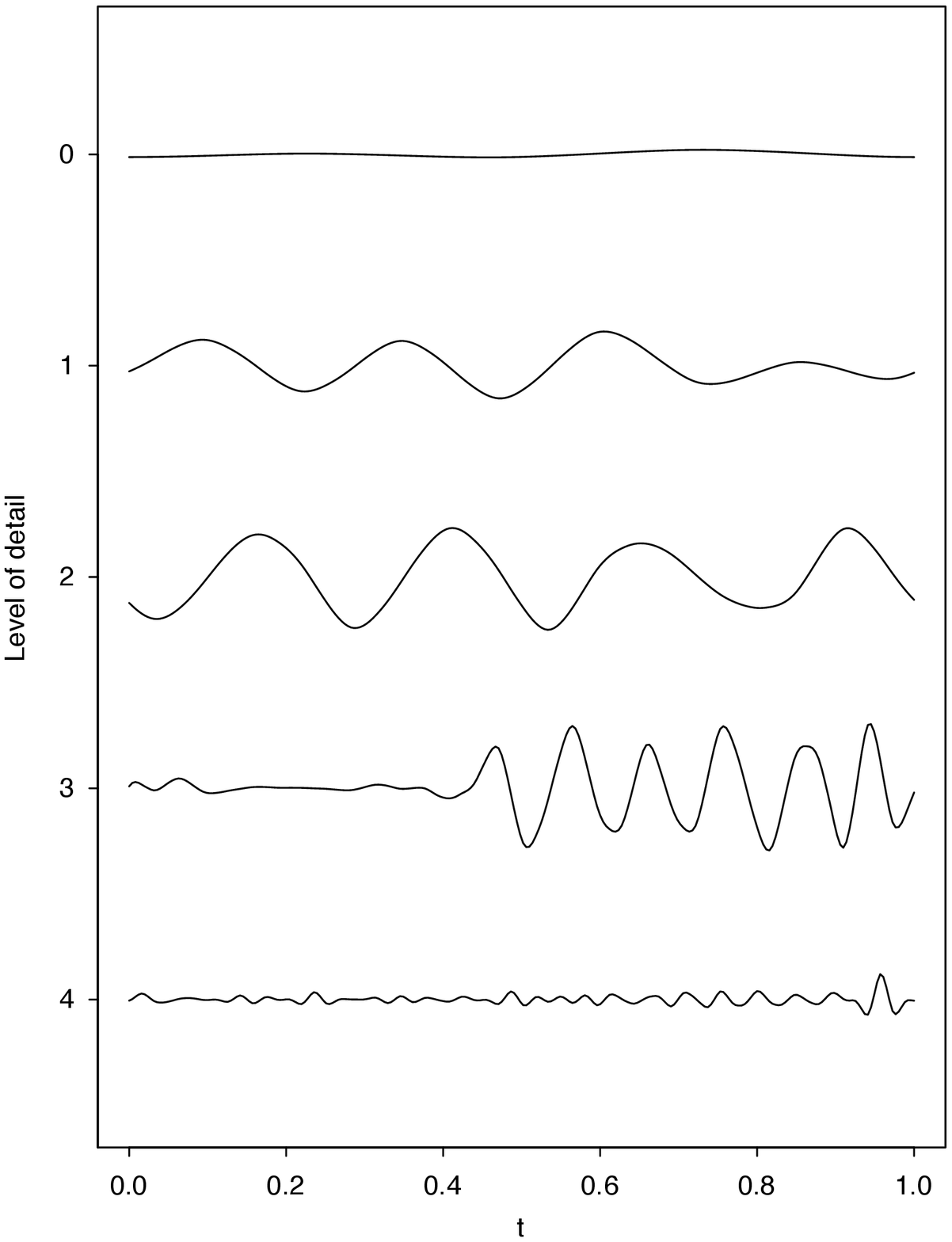} \end{tabular}
\caption{The left panel shows some simulated data which arises as a single sinusoidal term with frequency 4 when $t<0.5$, followed by a sum of sinusoidal terms with frequencies 4 and 10 when $t \ge 0.5$. The right panel shows the wavelet transform of these data which allows us to separate out the frequency behaviour at different levels of detail.}
\label{SimulatedData}
\end{figure}

To identify frequency behaviour from the wavelet output we can calculate the discrete wavelet transform. However, because the wavelets represent variation at different discrete frequencies, we cannot obtain exact frequency values for each species at a particular time. Instead, we can calculate approximate windows of the frequency value and show how these change over time. As an example, Figure \ref{SimulatedData} (left panel) shows a time series which exhibits seasonality at different scales over different time periods. The data are simulated from:
$$ y_t = \left\{ \begin{array} {ll} 
\beta_1 \sin(2\pi t f_1) + \epsilon_t & \mbox{for } t<0.5 \\
\beta_1 \sin(2\pi t f_1) + \beta_2 \sin(2 \pi t f_2) + \epsilon_t & \mbox{for } t\ge 0.5 \end{array}  \right.$$
where $\beta_1 = \beta_2 = 2$, $f_1=4$, $f_2=10$ and $\epsilon \sim N(0,0.3^2)$. The right panel of Figure \ref{SimulatedData} shows its associated wavelet transform using the Daubechies' Least Asymmetric basis. In comparison with Figure \ref{UnweightedWavelets} (which shows the unweighted wavelet bases) we can see the transition from one level of detail (level 2 to level 3 at approx $t=0.5$). The frequency values for these detail levels are 4 and 8 respectively, which are not identical to $f_1,f_2$ above but do give an approximation to the dynamic seasonality and allow us to determine the time at which frequencies might be changing.\\

\section{Modelling of multinomial data via nested ZaNI-binomial distributions}\label{multinomial}

The usual probability distribution used for constrained counts across multiple categories is the multinomial. However, this distribution can be restrictive when used with the common Dirichlet conjugate prior, as has long been known \citep{Aitchison1986}. Alternative link functions for the vector of proportions which aim to provide a richer set of behaviour for covariances between proportions have been proposed, for example the Isometric Log-Ratio \cite[ILR;][]{Egozcue2003,Parnell2013a} and the centralised log-ratio \citep[CLR; also known as the multinomial-logistic][]{Pawlowsky-Glahn2011}, though these too often have restrictions on the behaviour they can model and computational restrictions come into play when the number of categories is large. Other authors have avoided transformations and used richer distributions on the simplex \citep[e.g.][]{Pawlowsky-Glahn2011}. \\

An alternative approach, which remains computationally efficient even for large numbers of categories, is to decompose the count vector into nested partitions and use binomial (or binomial-like) distributions to model each section of the nesting independently. This decomposition may be identified either automatically or using informative prior information about the categories in question. In our case study we use the latter. This is related to the idea behind the nested Dirichlet distribution \citep[e.g.][]{Ng2008}. However, the nested Dirichlet does not take account of the zero and $N$ inflation that arises here.\\

We restrict our nesting structures to binary splits only, though multiple splits are also possible \citep{Salter-Townshend2012}. We write the nesting structure as $\mathcal{N} = \{ \mathcal{N}_1, \mathcal{N}_2, \ldots \}$ where each $\mathcal{N}_j$ defines a two-component partition of the data (or a subset of it) at that particular branch of the nesting structure. Note that the order of the nesting structure is irrelevant, since each sub-model is conditionally independent. An example nesting structure for our case study is shown in Figure \ref{nestingStructure}.\\

For a given nesting structure $\mathcal{N}$, vector of counts $y = \{y_1,\ldots,y_S\}$ with restriction $\sum_{i=1}^S y_i = K$ we write:
\begin{align}
\label{nestingEqn}
\pi(y|p) = \pi_{\mathcal{N}_1}\left(\sum_{j \in \mathcal{N}_1} y_j \; \bigg| \;p_{\mathcal{N}_1} \right) \times \pi_{\mathcal{N}_2}\left(\sum_{j \in \mathcal{N}_2} y_j\; \bigg| \;p_{\mathcal{N}_2}\right) \times \ldots
\end{align}
where $p_{\mathcal{N}_j}$ are the probabilities associated with that branch of the nesting structure. When fitted this way, each component of the product defines an individual likelihood which, when combined with suitable prior distributions, factorises over each branch of the nesting structure. Perhaps the obvious choice for $\pi$ above is the binomial, but we have found this to perform poorly due to zero inflation in one or both of the counts. To account for such inflation, we use a recently proposed zero and $N$-inflated Binomial distribution.\\

We consider each of our nested counts to arise from a zero inflated Poisson (ZIP) distribution \citep{Lambert1992} with some rate and zero inflation parameter. To simplify notation we consider $y$ and $z$ to arise from ZIP models, and $N=y+z$ to be their sum. We require the distribution $y|N$ which will naturally contain more 0 or $N$ values than expected by the standard binomial. A recently proposed distribution \citep{Sweeney2012} provides for a very neat formulation via a class of Zero and $N$ inflated (ZaNI) binomials.\\

Perhaps the first model for zero-inflated restricted sum data was that proposed by \cite{Hall2000}. Here a zero-inflated binomial was suggested of the form:
\begin{align}
y|N \sim \left\{ \begin{array}{lcl} 0 & \mbox{with probability} & 1-q_0 \\ Bin(N,p) & \mbox{with probability} & q_0 \end{array} \right.
\label{ZIbinomial}
\end{align}
where $q_0$ is the zero-inflation parameter. However, as noted by \cite{Sweeney2014}, this distribution has an inconsistency in that different inferences will obtained if $z$ and $y$ are swapped, and that it is inappropriate when both counts are zero-inflated. The specific parameterisation when $q_0$ is a constant also does not arise naturally from the underlying ZIP models. An elegant alternative, provided by \cite{Sweeney2012} sets:
\begin{align}
y|N \sim \left\{ \begin{array}{lcl} 0 & \mbox{with probability} & q_0 \\ N & \mbox{with probability} & q_N \\ Bin(N,p) & \mbox{with probability} & 1-q_0-q_N \end{array} \right.
\label{ZaNIbinomial}
\end{align}
where
$$q_0 = \frac{ e^{\lambda_0} (1-p)^{N} }{ e^{\lambda_0} (1-p)^N + e^{\lambda_N} p^N + 1 },\; q_N = \frac{ e^{\lambda_N} p^{N} }{ e^{\lambda_0} (1-p)^N + e^{\lambda_N} p^N + 1 }.$$
This distribution arises naturally from the constrained zero-inflated Poisson distribution and involves positive, real-valued hyper-parameters $\lambda_0$ and $\lambda_N$ which control zero and $N$ inflation respectively. Now when $p \rightarrow 0$, $q_0 \rightarrow \frac{1}{e^\lambda_0+1}$ and when $p \rightarrow 1$, $q_0 \rightarrow 1$, thus large values of $\lambda_0$ indicate high levels of zero inflation and similarly high values $\lambda_N$ indicate high levels of $N$ inflation. We further justify our choice of this model for our data in Section \ref{model}.\\

\section{Data}\label{data}

We use data from the Irish discards sampling programme which places observers onboard demersal fishing vessels to collect discard samples. Specifically, we use the number of fish from each species discarded per haul (number of hauls=680) from 1996 until the end of 2009 in the Irish Sea. Since the sampling targets are set by quarter, the data were collated to this time unit which spanned 56 quarters. \\

\begin{figure}[!h]
\begin{tabular}{cc} \includegraphics[width=0.5\textwidth]{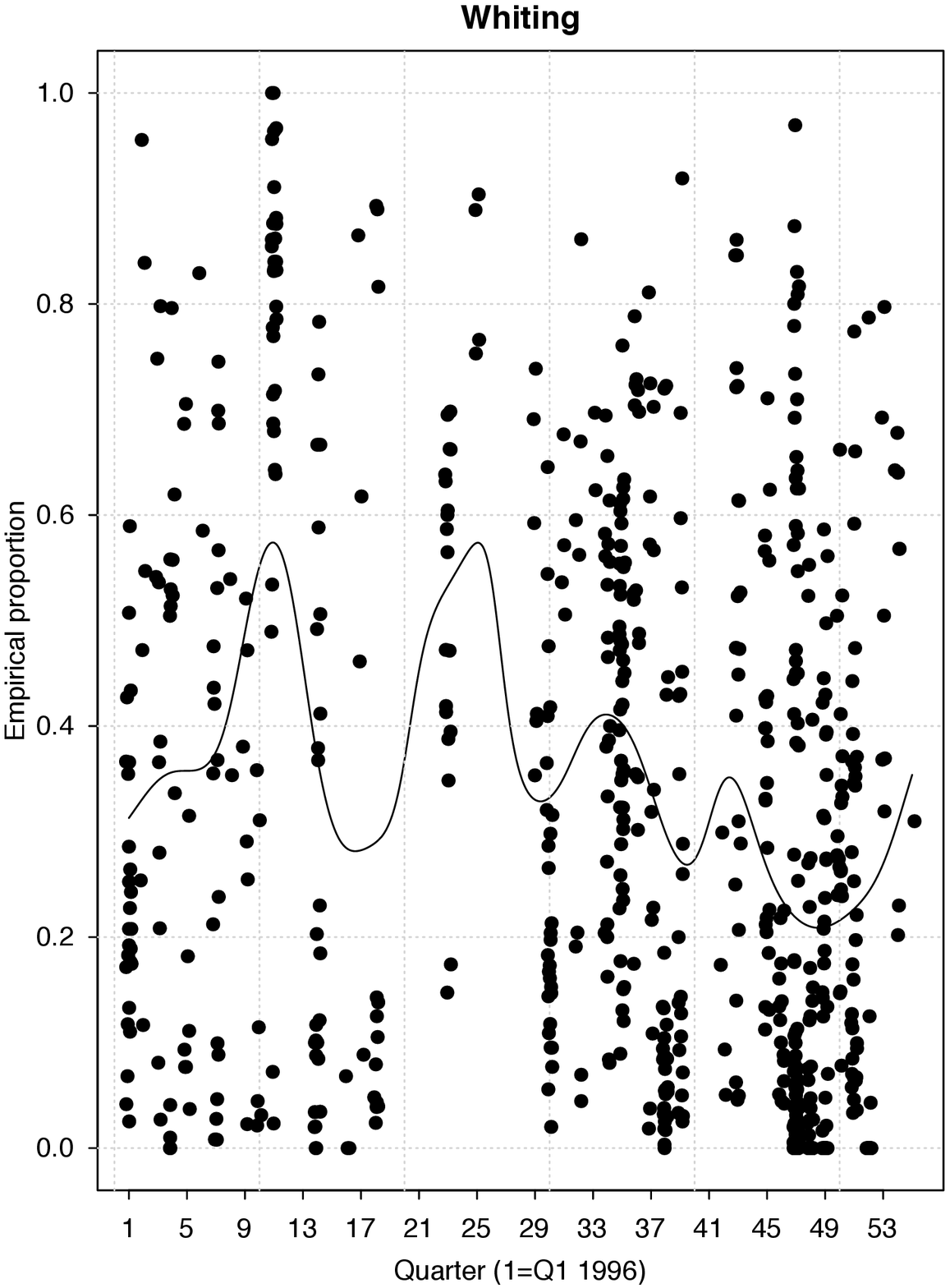} & \includegraphics[width=0.5\textwidth]{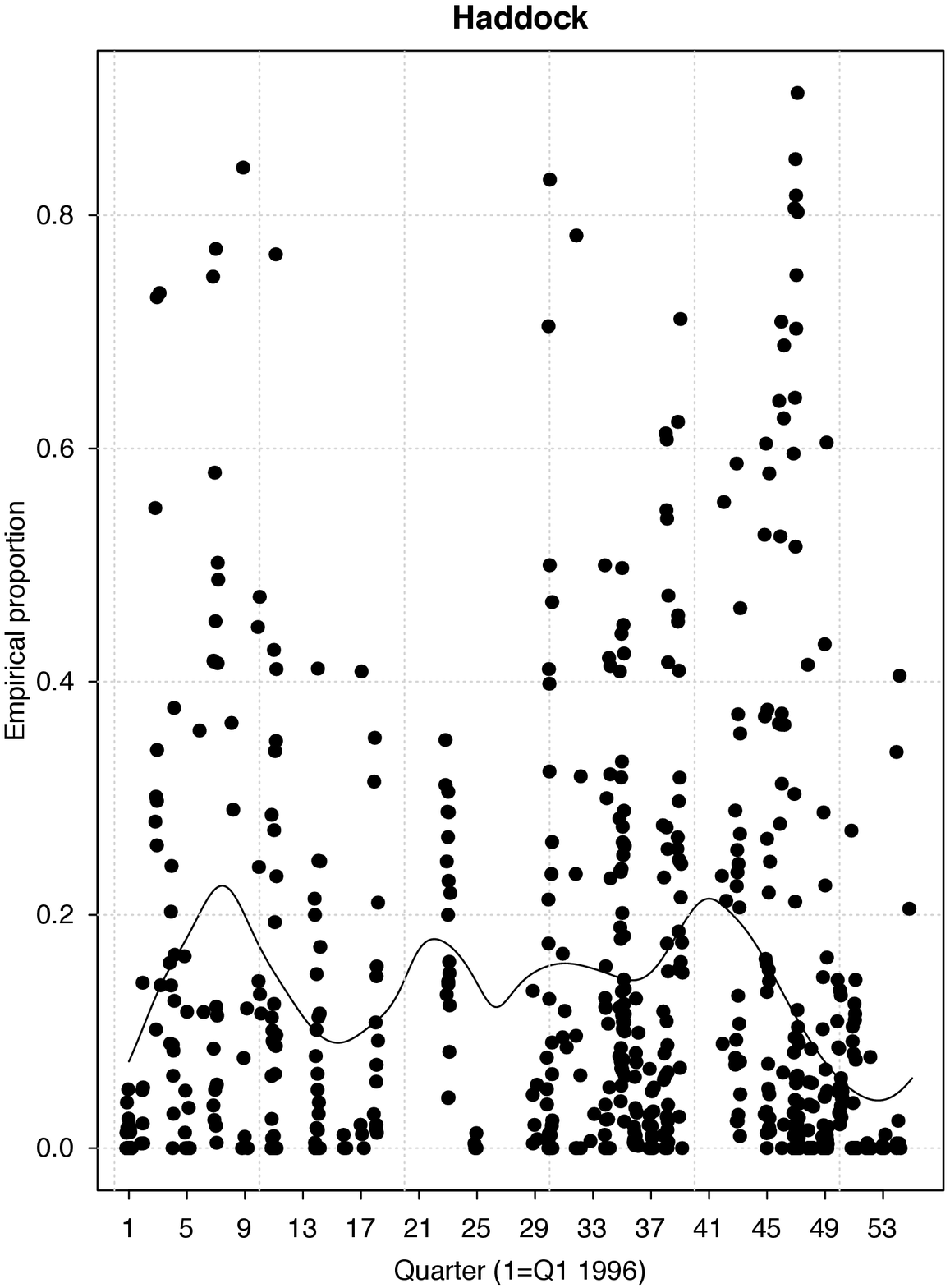} \end{tabular}
\caption{Plots of the empirical proportion for two of the most populous species over time. Left panel shows Whiting (\emph{Merlangius merlangus}); right panel shows Haddock (\emph{Melanogrammus aeglefinus}). The $x$-axis has been jittered slightly to enable clearer identification of individual points. The lines shown are kernel smoothers (bandwidth=5). }
\label{WhitingHaddock}
\end{figure}

Almost 80 species were discarded in the Irish Sea during the 14 year time-series investigated. For our model, we use seven commercial and non-commercial species. These are: whiting \emph{Merlangius merlangus}, haddock \emph{Melanogrammus aeglefinus}, dab \emph{Limanda limanda}, plaice \emph{Pleuronectes platessa}, grey gurnard \emph{Eutrigla gurnardus}, norway pout \emph{Trisopterus esmarkii} and poor cod \emph{Trisopterus minutus}. These seven constitute 85\% of all fish discarded in the Irish Sea. We aggregate the remaining 15\%, which also include some commercially important species such as cod (\emph{Gadus morhua}), under the label of `other species'. Plots of the empirical proportion ($y_j/N$) for the two most abundant species are shown in Figure \ref{WhitingHaddock}.\\

The temporal dynamics of discards depend on the characteristics of the fishery, the environment and the biology of the species. \cite{Viana2012} used univariate time series to show that there are annual cycles (i.e. 4 quarters) in cod, haddock and whiting discards of the Irish Sea likely owing to the seasonality in the spawning behaviour of the species. However, their model used a rigid Fourier basis. We therefore seek to determine whether these species are governed by similar dynamics in a more flexible modelling environment. We would like the model to identify these frequencies, whilst allowing for non-seasonal behaviour during certain time periods for certain species.\\

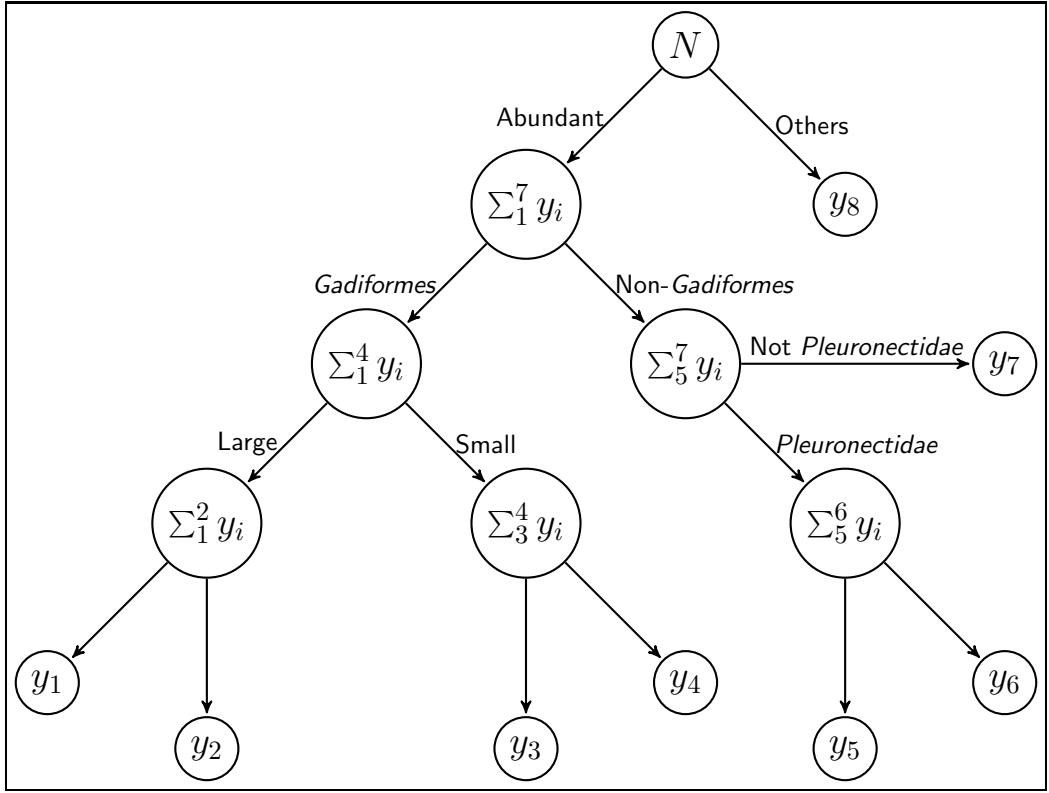
\begin{figure}[!p]
\fbox{\begin{tikzpicture}[->,>=stealth',shorten >=1pt,auto,node distance=3cm,
                    thick,main node/.style={circle,draw,font=\sffamily\Large\bfseries}]

  \node[main node] (1) {$N$};
  \node[main node] (2) [below left of=1] {$\sum_1^7 y_i$};
  \node[main node] (3) [below right of=1] {$y_8$};
  \node[main node] (4) [below left of=2] {$\sum_1^4 y_i$};
  \node[main node] (5) [below right of=2] {$\sum_5^7 y_i$};
  \node[main node] (6) [below left of=4] {$\sum_1^2 y_i$};
  \node[main node] (7) [below right of=4] {$\sum_3^4 y_i$};
  \node[main node] (8) [below left of=6] {$y_1$};
  \node[main node] (9) [below of=6] {$y_2$};
  \node[main node] (10) [below of = 7] {$y_3$};
  \node[main node] (11) [below right of=7] {$y_4$};
  \node[main node] (12) [below right of = 5] {$\sum_5^6 y_i$};
  \node[main node] (13) [below right of=3] {$y_7$};
  \node[main node] (14) [below of=12] {$y_5$};
  \node[main node] (15) [below right of=12] {$y_6$};

  \path[every node/.style={font=\sffamily\small}]
    (1) edge node [left] {Abundant} (2)
	    edge node [right] {Others} (3)
	(2) edge node [left] {\textit{Gadiformes}} (4)
	    edge node [right] {Non-\textit{Gadiformes}} (5)   
	(4) edge node [left] {Large} (6)
	    edge node [right] {Small} (7) 
	(6) edge node [left] {} (8)
	    edge node [right] {} (9)   
	(7) edge node [left] {} (10)
	    edge node [right] {} (11)  
	(5) edge node [right] {\textit{Pleuronectidae}} (12)
	    edge node [above] {Not \textit{Pleuronectidae}} (13)   
	(12) edge node [right] {} (14)
	    edge node [right] {} (15);   
    
\end{tikzpicture}}
\caption{The nesting structure we use to demonstrate the Wavelet ZaNI-binomial model. Here $y_1$=whiting, $y_2$=haddock, $y_3$=pout, $y_4$=poor cod, $y_5$=dab, $y_6$=plaice, $y_7$=grey gurnard, $y_8$=other species. Each internal node of the nesting structure defines an individual Wavelet ZaNI-binomial model.}
\label{nestingStructure}
\end{figure}

We use a nesting structure (Figure \ref{nestingStructure}) that takes into account the classification and size of the fish involved in our study. They are separated at the first level as being either `abundant' or `other species', then at lower levels according to being in the family\textit{Pleuronectidae} (right-eye flounders), or the order \textit{Gadiformes} (ray-finned). The nesting structure we use defines seven separate models which use the counts at various levels of aggregation as detailed in Figure \ref{nestingStructure}.\\

\begin{figure}[!h]
\includegraphics[width=0.9\textwidth]{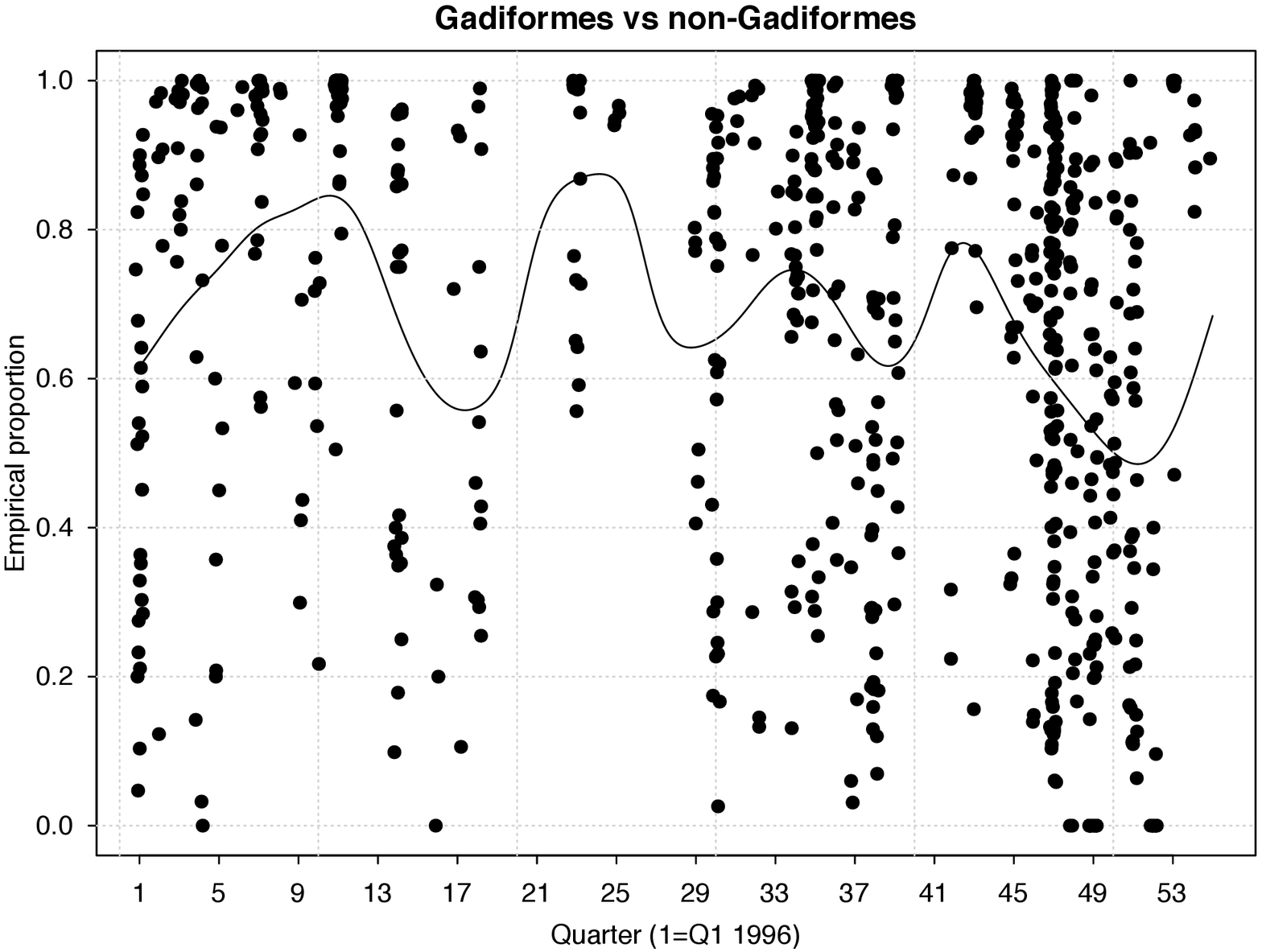}
\caption{A plot of the empirical proportion for some aggregated data corresponding to \textit{Gadiformes} vs non-\textit{Gadiformes} - see the nesting structure of Figure \ref{nestingStructure}. The line shows a kernel smoother (bandwidth=5). Approximately 10\% of the data are either zero or $N$ (corresponding here to $\hat{p}=1$.}
\label{Gadiformes}
\end{figure}

As we progress up the nesting structure, the likelihood of zero and $N$ inflation increases. A plot of the comparison of the aggregate \textit{Gadiformes} vs \textit{non-Gadiformes} (Figure \ref{Gadiformes}) shows that 10\% of the data are either $N$ or zero. This is because species 1 and 2 (haddock and whiting) are the most abundant of all species and thus take up all of the proportion in many cases. We might expect values of $\lambda_0$ or $\lambda_N$ for these models to be high.\\

\section{Bayesian Models}\label{model}

In this section we outline the notation and model structure for five competing models which we later compare using WAIC \citep{Watanabe2013} and out-of-sample evaluation. The models we compare are:
\begin{enumerate}
\item A multinomial wavelet model 
\item A nested binomial model with constant mean
\item A nested binomial wavelet model
\item A nested ZI-binomial wavelet model
\item A nested ZaNI-binomial wavelet model
\end{enumerate}
The wavelet structures used are those outlined in Section \ref{wavelets}. The nesting structure used is that defined in Figure \ref{nestingStructure}. We include a binomial with constant mean to detect for possible Slutsky effects \citep{Slutzky1937} where irregular seasonality can be spuriously obtained from smoothing.
Below we provide notation for the most general case (multinomial wavelet) which can then be simplified for the other three models.

\subsection{Notation}
\begin{itemize}
\item $y_{ijk}(t)$ is a count of fish species $k=1,\ldots,K$ on trip $j=1,\ldots,J$, observation $i=1,\ldots,n_j$ at time $t=1,\ldots,T$. $y_{ij}$ is the $K$-vector of counts for species $k$. In total we have $\sum_{j=1}^J n_j = M$ counts for each species.
\item $N_{ij}(t)= \sum_{k=1}^8 y_{ijk}(t)$ is the total number of all fish species on trip $j$, observation $i$ at time $t$.
\item $\lambda_{0k},\lambda_{Nk}$ are zero/$N$ inflation parameters associated with species $k$ respectively.
\item $p_{ijk}(t)$ is the proportion of fish species $k$ on trip $j$, observation $i$ at time $t$.
\item $\mu_k(t)$ is the mean proportion (logit scale) for species $k$ at  time $t$. $\mu_k$ is a $T$-vector of means for species $k$.
\item $\boldmath{B}$ is the $M\times T$ indicator matrix which maps each mean $\mu_k(t)$ on to a time point $t$ associated with $y_{ijk}(t)$.
\item $b_{jk}$ is the random effect for trip $j$ on species $k$ with associated variance $\sigma^2_{uk}$.
\item $\sigma^2_k$ is the over-dispersion variance associated with proportion $p_{ijk}(t)$
\item $\boldmath{H}$ is a spline interpolating matrix of dimension $T$ by $L$ where $L=2^D$ is the number of wavelet basis functions and $D$ is the number of wavelet detail levels.
\item $\boldmath{W}$ is a wavelet basis function matrix of dimension $L$ by $L$.
\item $\theta_{kl}$ is the weight coefficient associated with species $k$ and wavelet basis function $l=1,\ldots,L$. $\theta_k$ is the $L$-vector of wavelet basis function coefficients.
\item $\phi_k$ is a global shrinkage parameter for $\theta_k$.
\item $\tau_{d(l),k}$ is a cumulative local shrinkage parameter associated with the detail of wavelet basis function $l$, denoted $d(l)$, for species $k$.
\item $\delta_{d(l),k}$ is a local shrinkage parameter associated with the detail of wavelet basis function $l$ and species $k$.
\item $\alpha_{1k},\alpha_{2k}$ are hyper-parameters associated with the local shrinkage.
\item $\nu_k$ is a hyper-parameter associated with global shrinkage. We fix this value.
\end{itemize}
In general we write, e.g. $y$, to denote the entire collection of $y_{ijk}(t)$ values, and similarly with other parameters. Below we outline the model structure which uses these parameters.

\subsection{Model structure}

Our first model sets $y_{ij}|N_{ij},p_{ij}$ as a multinomial-logistic distribution, and thus involves no nesting structure. The goal is to find the posterior distribution of parameters $\Theta = \{p, \mu, b, \sigma^2_u, \sigma^2, \theta, \phi, \tau, \delta, \alpha_1, \alpha_2, \nu\}$ given data $y$ and $N$. Of main interest are parameters $p$ (the species proportions), $\mu$ (the temporal mean dynamics), and $\theta$ (the wavelet coefficients). Written out in full, we have:
\begin{align}
\label{bigEquation}
\pi(\Theta|y,N) \propto&\; \pi(y|N,p) \pi(p|\mu,b,\sigma^2_u,\sigma^2) \pi(\mu|\theta) \pi(\theta|\tau,\phi) \pi(\phi|\nu) \pi(\tau|\delta) \\ 
& \; \pi(\delta|\alpha_1,\alpha_2) \pi(\alpha_1,\alpha_2,\nu) \nonumber \\
\propto&  \left[ \prod_{j=1}^J \prod_{i=1}^{n_j} \pi(y_{ij}|N_{ij},p_{ij}) \right] \left[ \prod_{k=1}^K \prod_{j=1}^J \prod_{i=1}^{n_j} \pi(p_{ijk}|\mu,b_j,\sigma_k^2) \right] \nonumber \\
& \left[ \prod_{k=1}^K \pi(\mu_k|\theta_k) \right] \left[\prod_{k=1}^K \prod_{j=1}^J \pi(b_{jk}|\sigma^2_{uk}) \right] \left[ \prod_{k=1}^K \prod_{l=1}^L \pi(\theta_{lk}|\phi_k,\tau_{d(l),k}) \right] \left[ \prod_{k=1}^K \pi(\tau_{d(l)},k|\delta_{d(l),k})  \right] \nonumber  \\
& \left[ \prod_{k=1}^K \pi(\phi_k|\nu) \pi(\delta_{d(l),k}|\alpha_{1k},\alpha_{2k}) \pi(\alpha_{1k},\alpha_{2k},\nu) \nonumber \right]
\end{align} 
The individual distributions are:
\begin{eqnarray*}
y_{ij} &\sim&  \mathrm{Multinomial}(N_{ij},p_{ij}),\\
\mathrm{multilogit}(p_{ijk}) &\sim& N(B_{ij}\mu_k + b_{jk} ,\sigma^2_k),\\
\mu &=& H W^t\theta,\\
b_{jk} &\sim& N(0,\sigma_{uk}^2),\\
\theta &\sim& N\left(0,(\phi \tau_{d(l)})^{-1}\right),\\
\tau_{d(l),k}|\delta_{d(l),k} &=& \prod_{s=1}^{d(l)} \delta_{sk}\\
\delta_{1k} &\sim& Ga(\alpha_{1k},1),\\
\delta_{d(l)>1,k} &\sim& Ga(\alpha_{2k},1)
\end{eqnarray*}
We outline prior distributions for the remaining terms below. \\

For the nested binomial, ZI-binomial and ZaNI-binomial, we introduce the given nesting structure $\mathcal{N}$ which defines pairs $\{\tilde{y}_{ij1},\tilde{N}_{ij1}\},\; \{\tilde{y}_{ij2},\tilde{N}_{ij2}\}, \ldots$ formed as aggregates of the individual counts. For example, for the  second layer of the nesting structure given in Figure \ref{nestingStructure} split by \textit{Gadiformes} vs non-\textit{Gadiformes}, $\tilde{y}_{ij2} = \sum_{k=1}^4 y_{ijk}$ and $\tilde{N}_{ij2} = \sum_{k=1}^7 y_{ijk}$. Under this formulation, Equation \ref{bigEquation} factorises over each nesting partition $k$ and thus each pair becomes a separate model. This greatly speeds up model fitting, and we now use the logit transformation rather than the multinomial-logit.\\

For the binomial models at nesting level $k$ we have:
$$\tilde{y}_{ijk} \sim \mathrm{binomial}(\tilde{N}_{ijk},p_{ij}).$$
For the constant mean binomial version we set all $\theta_l=0$ so only the random effect and over-dispersion parameters remain. For the ZI-binomial
$$\tilde{y}_{ijk} \sim \mbox{ZI-binomial}(\tilde{N}_{ijk},p_{ij},q_0),$$
where ZI-binomial is defined in Equation \ref{ZIbinomial} and $q_0 = \frac{1}{e^{\lambda_0}(1-p)^N+1}$. Finally, for the ZaNI-binomial we have:
$$\tilde{y}_{ijk} \sim \mbox{ZaNI-binomial}(\tilde{N}_{ijk},p_{ij},q_0,q_N),$$
where the ZaNI-binomial is as defined in Equation \ref{ZaNIbinomial} with $q_0 = \frac{ e^{\lambda_0} (1-p)^{N} }{ e^{\lambda_0} (1-p)^N + e^{\lambda_N} p^N + 1 }$ and $q_N = \frac{ e^{\lambda_N} p^{N} }{ e^{\lambda_0} (1-p)^N + e^{\lambda_N} p^N + 1 }$. Aside from the factorisation over $k$ and the addition of prior distributions for $\lambda_0$ and $\lambda_N$, the posterior is identical to Equation \ref{bigEquation}.\\

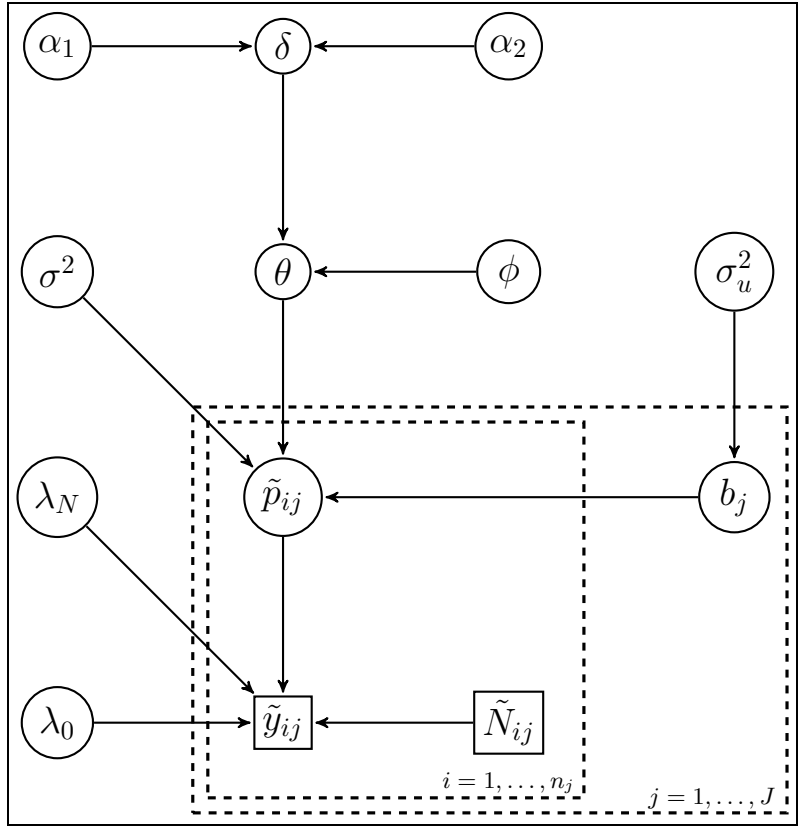
\begin{figure}[!h]
\fbox{\begin{tikzpicture}[->,>=stealth',shorten >=1pt,auto,node distance=3cm,
                    thick,main node/.style={circle,draw,font=\sffamily\Large\bfseries}]

  \draw[dashed, very thick] (-1,-1)rectangle (4,4);
  \draw[dashed, very thick] (-1.2,-1.2)rectangle (6.7,4.2);
  \node[] at (3,-0.8) {$i=1,\ldots,n_j$};	
  \node[] at (5.7,-1) {$j=1,\ldots,J$};	
  
  \node[rectangle,draw,font=\sffamily\Large\bfseries] (1) {$\tilde{y}_{ij}$};
  \node[rectangle,draw,font=\sffamily\Large\bfseries] (2) [right of=1] {$\tilde{N}_{ij}$};
  \node[main node] (3) [above of=1] {$\tilde{p}_{ij}$};
  \node[main node] (4) [above of=3] {$\theta$};
  \node[main node] (5) [above of=4] {$\delta$};
  \node[main node] (6) [right of=4] {$\phi$};
  \node[main node] (7) [left of=5] {$\alpha_1$};
  \node[main node] (8) [right of=5] {$\alpha_2$};
  \node[main node] (9) [left of=1] {$\lambda_0$};
  \node[main node] (10) [left of=3] {$\lambda_N$};
  \node[main node] (11) [left of=4] {$\sigma^2$};
  \node[main node] (12) [right of=6] {$\sigma^2_u$};
  \node[main node] (13) [below of = 12] {$b_j$};

  \path[every node/.style={font=\sffamily\small}]
	(2) edge node [left] {} (1)
	(3) edge node [left] {} (1)
	(4) edge node [left] {} (3)
	(5) edge node [left] {} (4)	
	(6) edge node [left] {} (4)	
	(7) edge node [left] {} (5)	
	(8) edge node [left] {} (5)	
	(9) edge node [left] {} (1)	
	(10) edge node [left] {} (1)	
	(11) edge node [left] {} (3)	
	(12) edge node [left] {} (13)	
	(13) edge node [left] {} (3);  
    
\end{tikzpicture}}
\caption{A Directed Acyclic Graph of the ZaNI-binomial model we fit to our fishery discards data. The binomial model is equivalent but removes nodes $\lambda_0$ and $\lambda_N$, whilst the ZI-binomial model is equivalent but removes only $\lambda_N$. The constant mean binomial model also removes $\theta$ and all its ancestral nodes.}
\label{DAG}
\end{figure}

We place weakly informative prior distributions on the remaining hyper-parameters, with the exception of $\nu$, for which we follow \cite{Bhattacharya2011} and set at $\nu=3$. Experimentation indicates that the model is not sensitive to this setting. Overall:
\begin{itemize}
\item $\phi_k|\nu \sim Ga(\nu/2,\nu/2)$
\item $\nu=3$
\item $\alpha_{1k} \sim U(0,50)$
\item $\alpha_{2k} \sim U(1,50)$
\item $\sigma_{uk} \sim |t|_1(0,100)$
\item $\sigma_k \sim |t|_1(0,100)$
\item $\lambda_{0k},\lambda_{Nk} \sim N(0,100)$
\end{itemize}
A DAG for our ZaNI-binomial model is shown in Figure \ref{DAG}.\\

\section{Model fitting and results}\label{results}

Prior to fitting, we wrap the edges of the data to avoid edge effects. All models are fitted using STAN v2.3 \citep{Team2014} using the No-U-Turn sampler \citep[NUTS, ][]{Hoffman2014}, a variant of Hamiltonian Monte Carlo in which all parameters are updated simultaneously using the gradient information in the un-normalised posterior. The idea behind NUTS is that it avoids the need to tune the number of steps and the step size, which can often cause deterioration of the model output.\\

Using STAN, we run each of the models for 2000 iterations (1000 burn-in, no thinning) with 3 chains and check convergence using the R-hat measure \cite{brooksgelman1,gelmanrubin1}. All of our model runs converged after this length of time. All models were fitted on a 2.6GHz Core i7 processor with 16Gb RAM. The multinomial model was slower, taking $\approx$ 2 hours to run. The binomials models however, could be fitted in $\approx$ 20 minutes for all runs of the nesting level. This could be further reduced through parallelisation, though we do not use this here as we found convergence was improved when using the values of the simpler model run as starting values for a more complex model.\\

For each model we compute the Wanatabe Akaike Information Criterion \cite[WAIC; ][]{Watanabe2013}. Like other information criteria, the WAIC aims to estimate the degree to which the model fit underestimates model error. The WAIC differs from others in that it is fully Bayesian unlike, say, the deviance information criterion. In particular the WAIC penalises the deviance by a measure based on the sum of the variances of the log density, which is estimated from the posterior samples. We use:
$$\mathrm{WAIC} = -2\widehat{\mathrm{lpd}} + 2p_{\mathrm{WAIC}}$$
where $\widehat{\mathrm{lpd}}$ is the estimated log predictive density, defined as 
$$\widehat{\mathrm{lpd}} = \sum_{i=1}^M \log \left[ \frac{1}{H} \sum_{h=1}^H p(y_i|\Theta^h) \right]$$
where $H$ denotes the total number of posterior samples and $\Theta^h$ denotes a posterior draw. The penalty $p_{\mathrm{WAIC}}$ is a measure of the effective number of parameters and is defined as the sum of the sample variances of the log density across the parameters:
$$p_{\mathrm{WAIC}} = \sum_{i=1}^M \frac{1}{H-1} \sum_{h=1}^H \left( \log p(y_i|\Theta^h) - \overline{\log p(y_i|\Theta)} \right)^2$$
where $\overline{\log p(y_i|\Theta)}$ indicates the mean with respect to the posterior samples. This penalty tends to be more stable than that suggested by DIC and, by definition, guarantees a positive value for the effective number of parameters.\\

As an alternative measure of model fit, we also create a 10\% holdout data set and test predictive performance of the models. Despite WAIC also aiming to estimate the out-of-sample fit of these models, we follow \citep{Vehtariy2014} in our belief that these answer differing questions about the models based on both finite and asymptotic behaviour. To make things slightly simpler, the hold-out set we create is only sampled from time points where we have multiple observations, as this means that $\mu_t$ will have already been created as part of the model run.\\

We first consider model comparison via WAIC. We fit the wavelet multinomial-logit, nested constant mean binomial, nested wavelet binomial, nested wavelet ZI-binomial, and nested wavelet ZaNI-binomial model to the full data set, and create WAIC values using code provided in \citep{Vehtariy2014}. Since WAIC is calculated on the deviance scale, we can add values provided by each nested-model and compare both multinomial and binomial models. The values are shown in Table \ref{WAIC_table}. We can conclude from the totals that the nesting and the zero and $N$ inflation are beneficial over the multinomial. Furthermore, since all the binomial models are fitted on the same nesting structure we can look at the WAIC for each nested model. We note that the ZaNI-binomial is competitive for all models and seems more appropriate at more aggregated levels of the nesting structure. This is to be expected as zero and $N$ inflation is likely to be higher here.\\

\begin{table}[!h]
\begin{tabular}{l|rrrrr}
\hline
\hline
Nested Model & CM-B & W-B & W-ZI-B & W-ZaNI-B & Multinomial\\
\hline
Whiting vs Haddock & 6181 & 6172 & 6177 & \textbf{6158} & -\\
\hline
Pout vs Poor cod & \textbf{2865} & \textbf{2865} & 2877 & 2890 & -\\
\hline
Dab vs Plaice & 4172 & \textbf{4169} & 4185 & 4205 & -\\
\hline
Large vs small & 6549 & 6549 & 6533 & \textbf{6456} & -\\
\hline
\textit{Pleuronectidae} vs non-\textit{Pleuronectidae} & 5609 & 5620 & 5616 & \textbf{5600} & -\\
\hline
\textit{Gadiformes} vs non-\textit{Gadiformes} & 8266 & 8273 & 8241 & \textbf{8203} & -\\
\hline
Abundant vs Others & 8944 & 8950 & 8948 & \textbf{8913} & -\\
\hline
Total & 42586 & 42598 & 42577 & \textbf{42425} & 42960 \\
\hline 
\hline
\end{tabular}
\caption{WAIC values for the different nested models attempted at the different levels of the nesting structure. Lower values indicate a better model. The smallest values for each model are shown in bold. The models are: CM-B = constant mean binomial, W-B = Wavelet binomial, W-ZI-B = Wavelet ZI-binomial, W-ZaNI-B = Wavelet ZaNI binomial, Multinomial = Wavelet multinomial. Note that the multinomial model is not nested and thus has no individual components} 
\label{WAIC_table}
\end{table}

\begin{figure}[!p]
\begin{tabular}{cc} 
\includegraphics[width=0.5\textwidth]{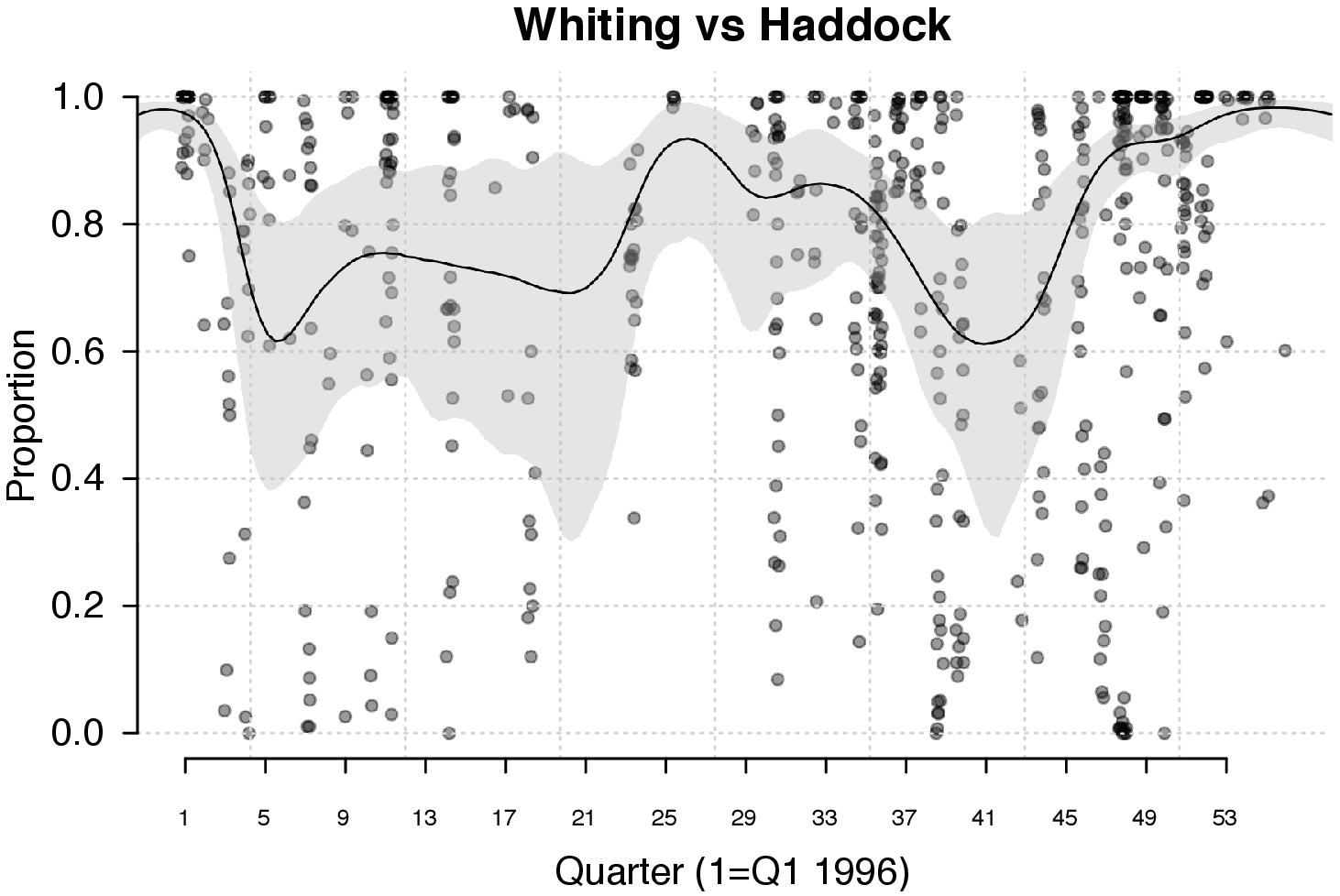} & \includegraphics[width=0.5\textwidth]{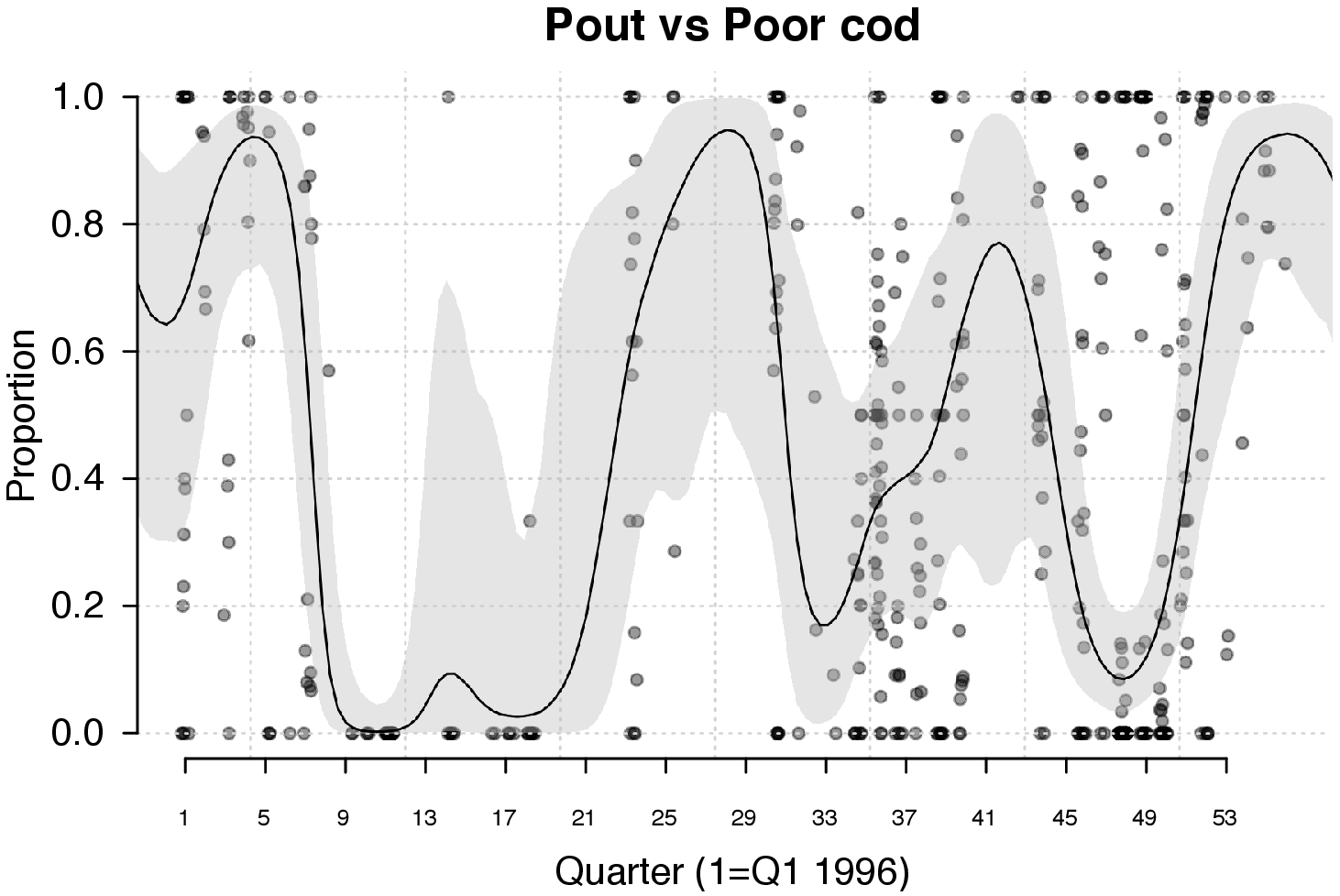} \\
\includegraphics[width=0.5\textwidth]{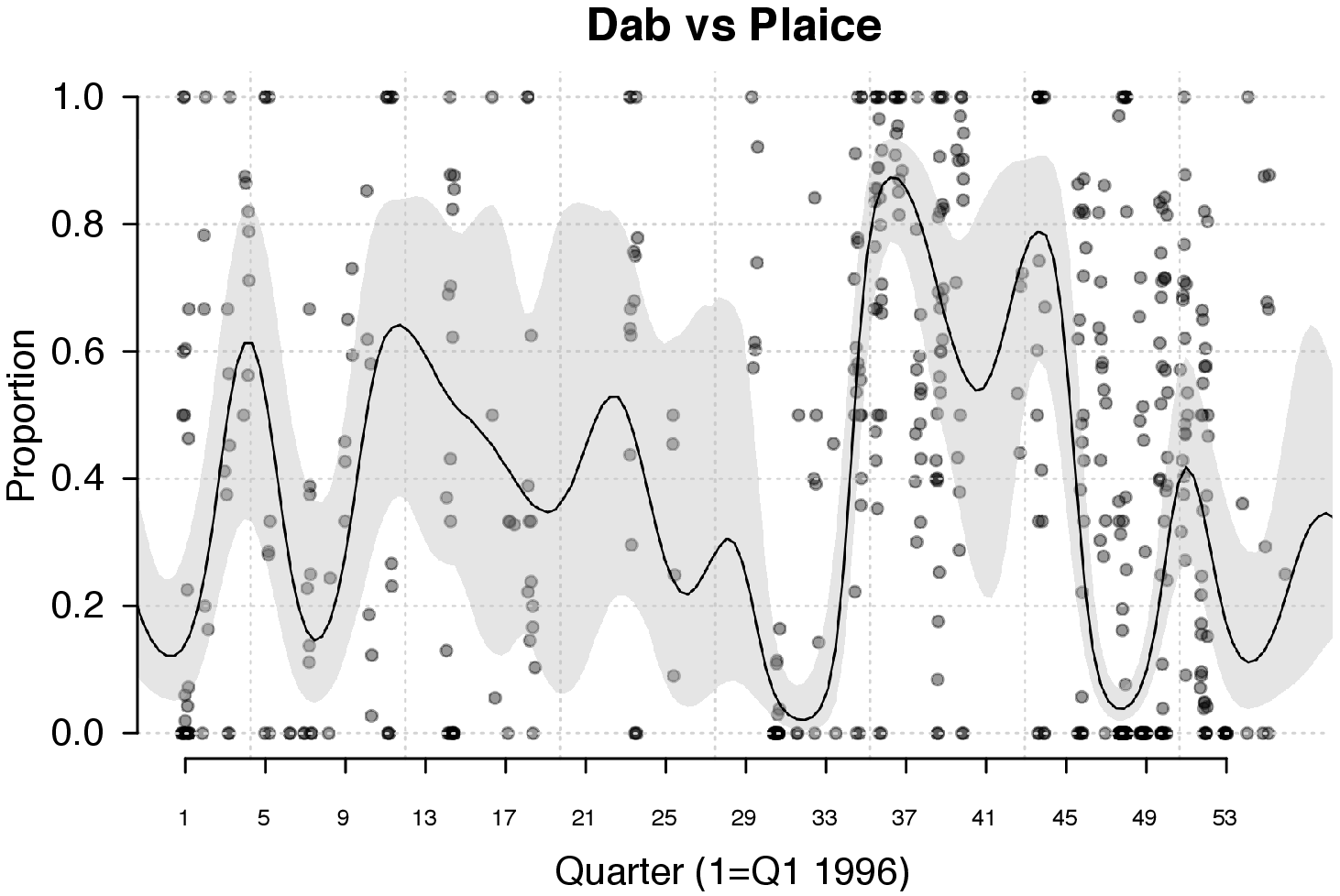} & \includegraphics[width=0.5\textwidth]{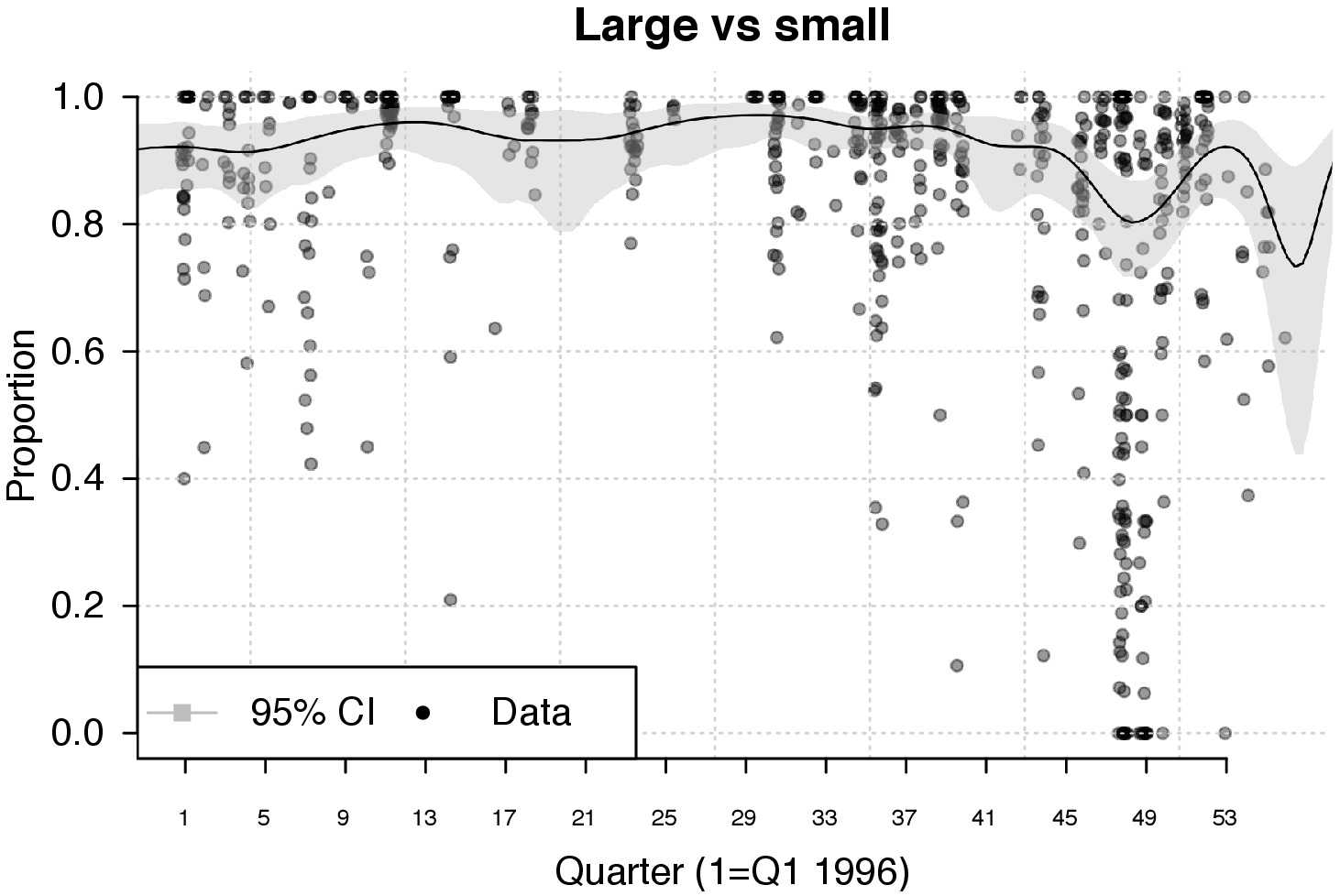} \\
\includegraphics[width=0.5\textwidth]{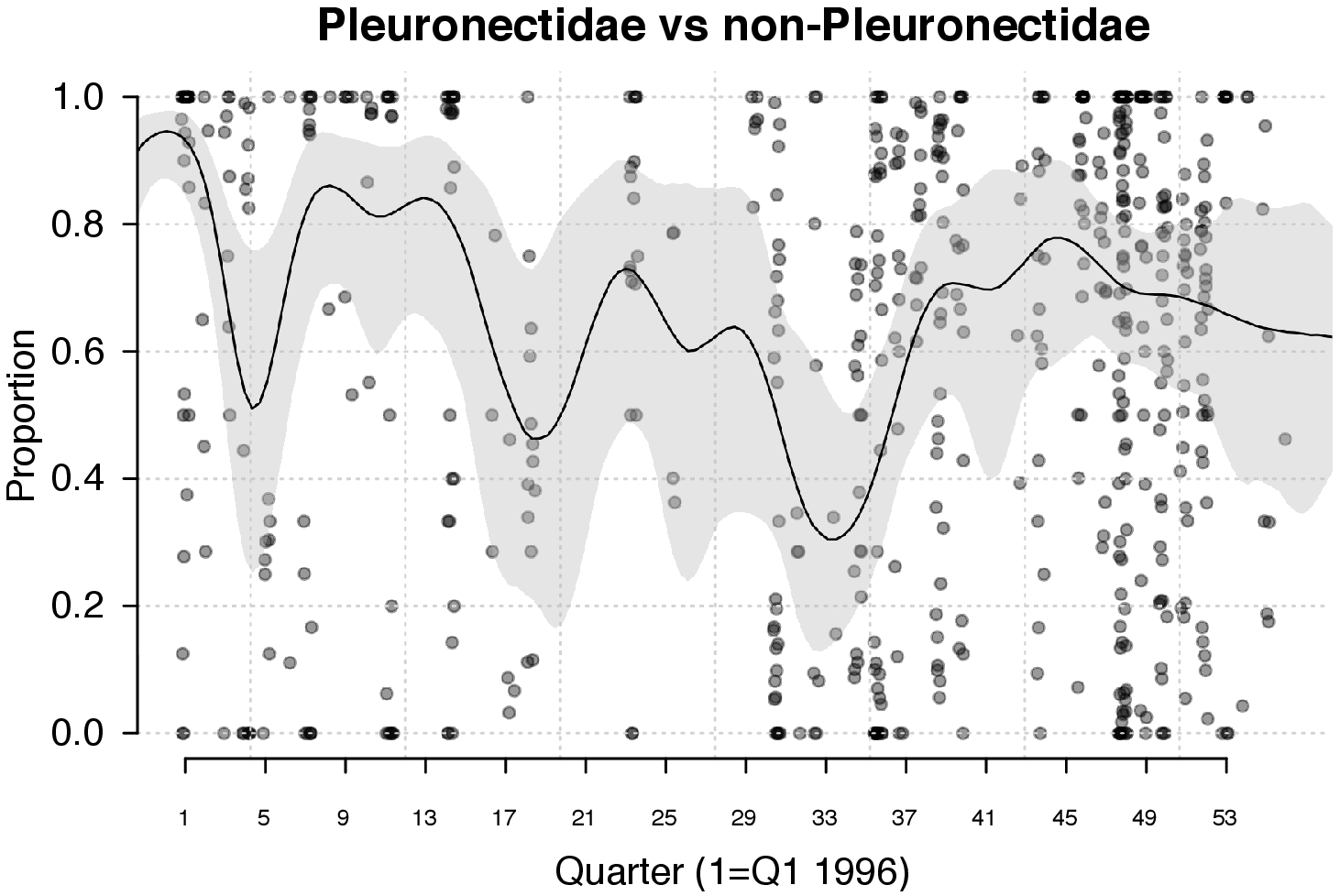} & \includegraphics[width=0.5\textwidth]{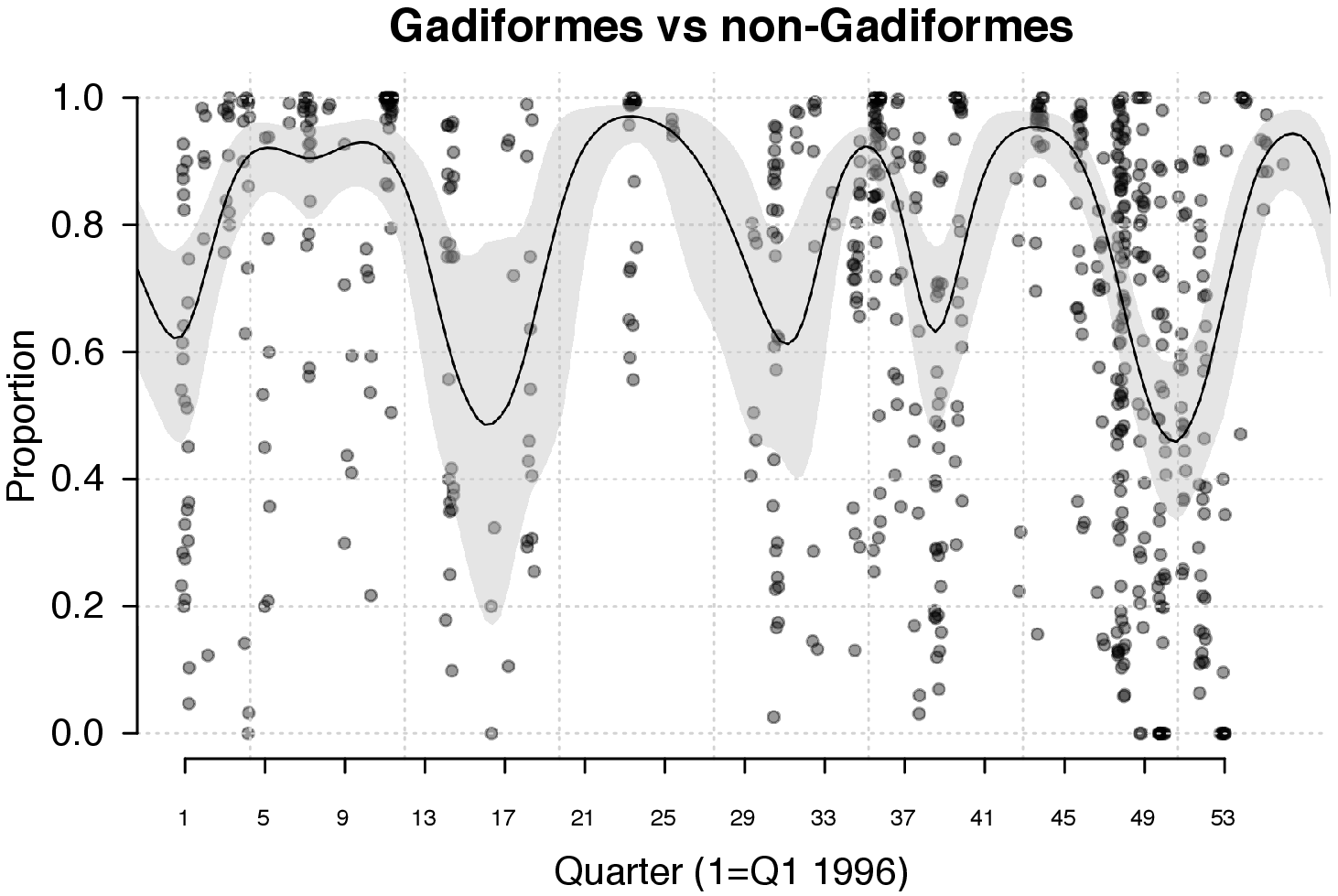} \\
\includegraphics[width=0.5\textwidth]{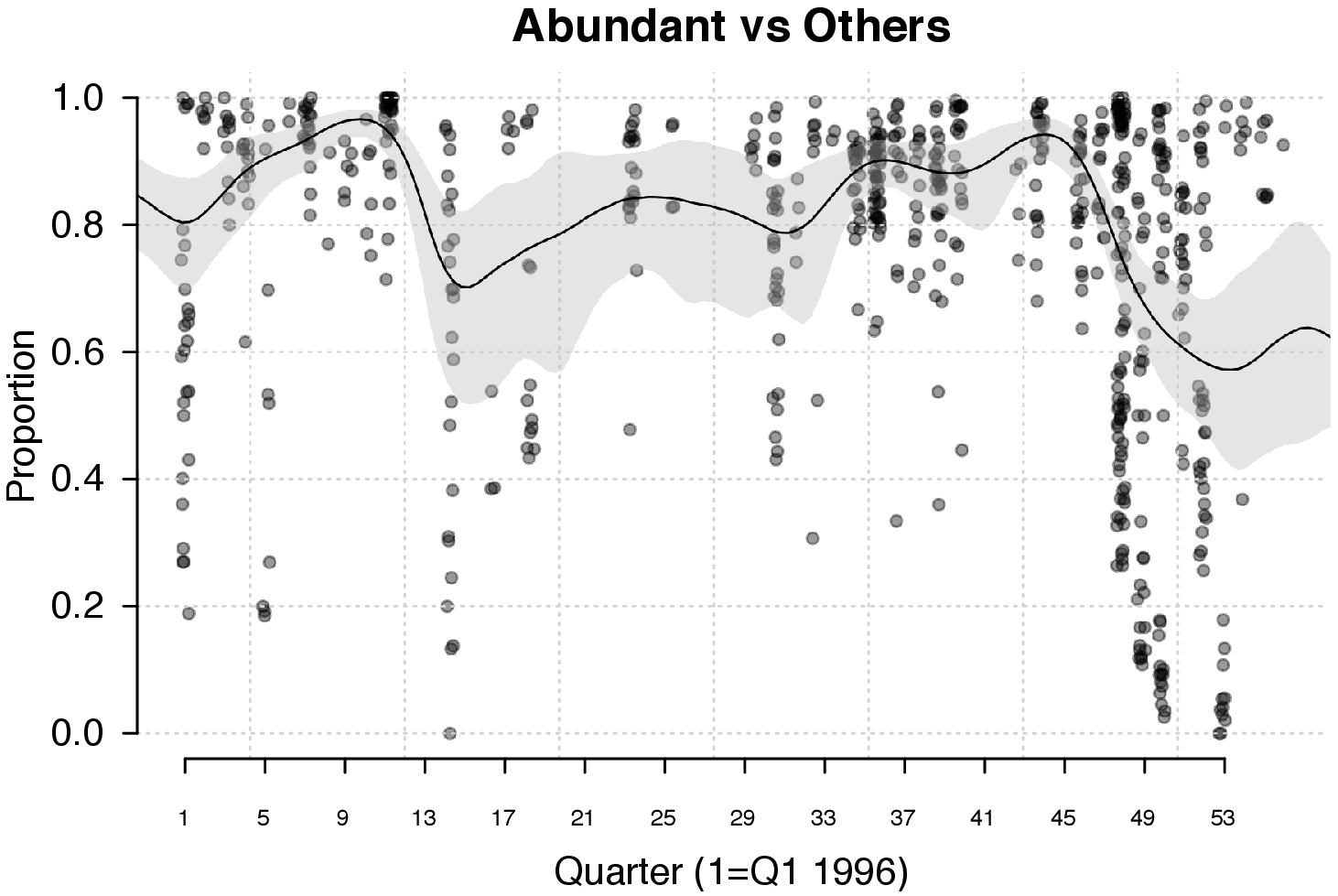} &  \\
\end{tabular}
\caption{Sample proportions and fitted medians (with shaded 95\% credible intervals) for the 7 different models defined by the nesting structure of Figure \ref{nestingStructure}. In each plot the $x$-axis is jittered slightly so that individual data points can be seen more clearly.}
\label{all_fits}
\end{figure}

\begin{figure}[!p]
\begin{tabular}{cc} 
\includegraphics[width=0.5\textwidth]{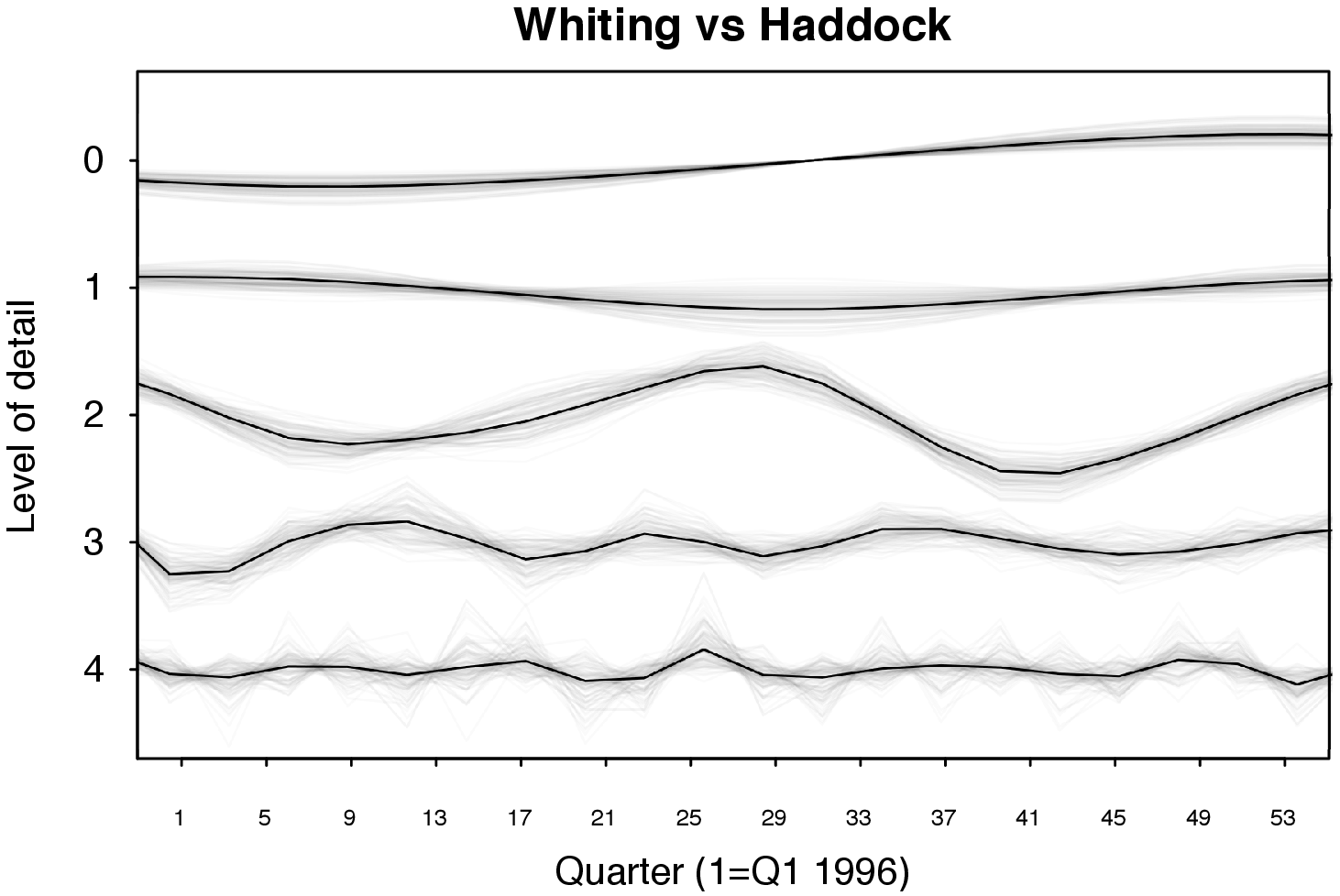} & \includegraphics[width=0.5\textwidth]{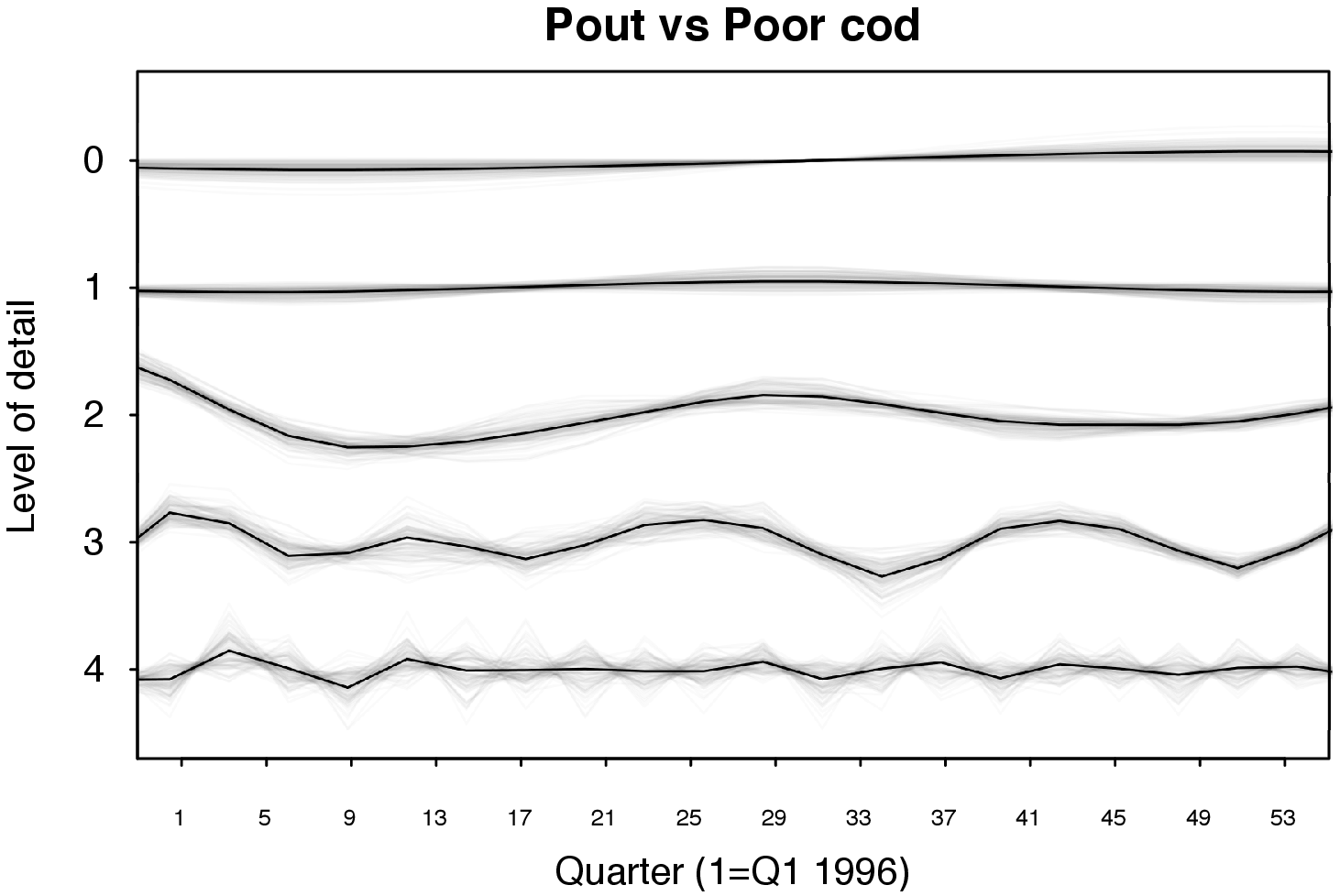} \\
\includegraphics[width=0.5\textwidth]{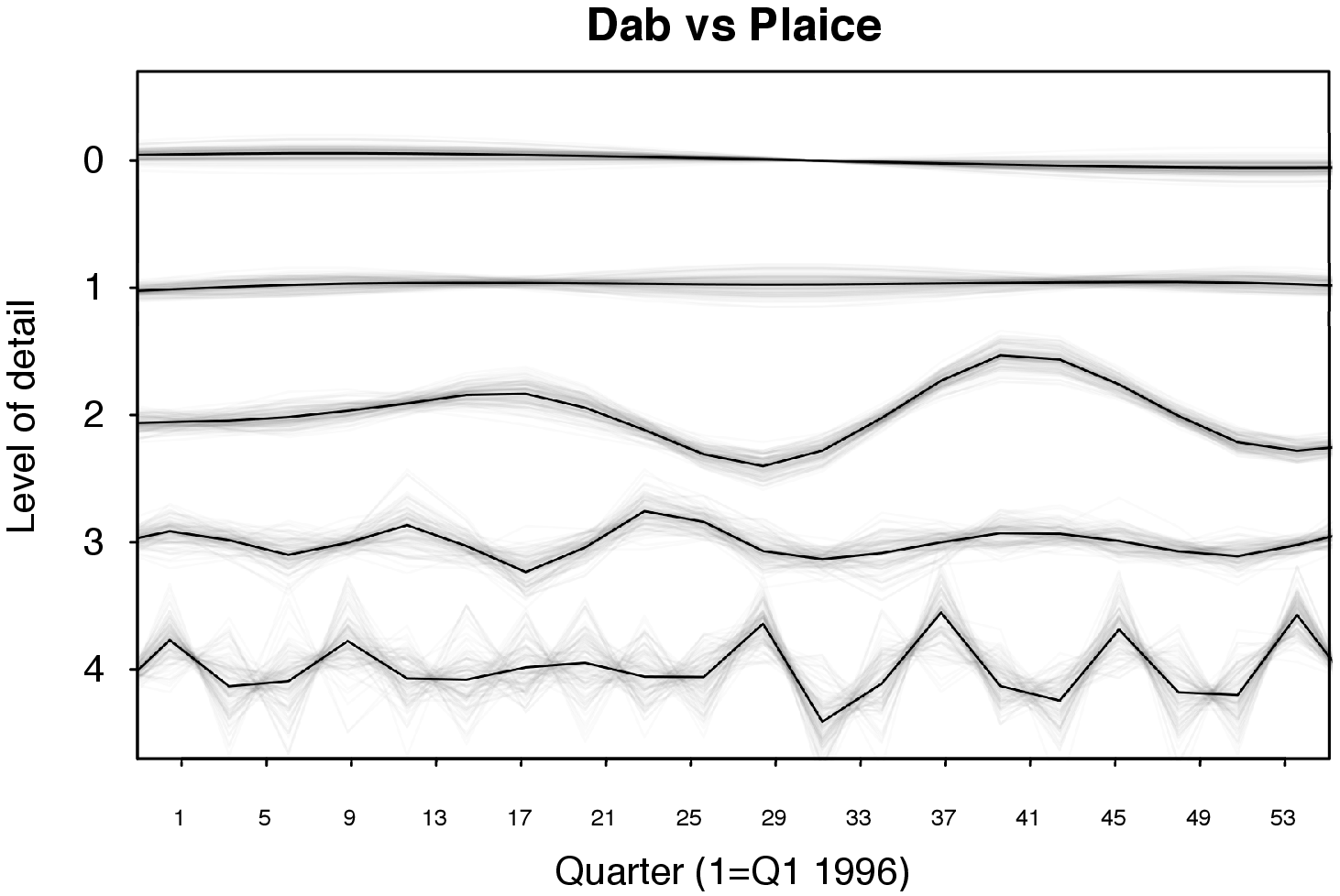} & \includegraphics[width=0.5\textwidth]{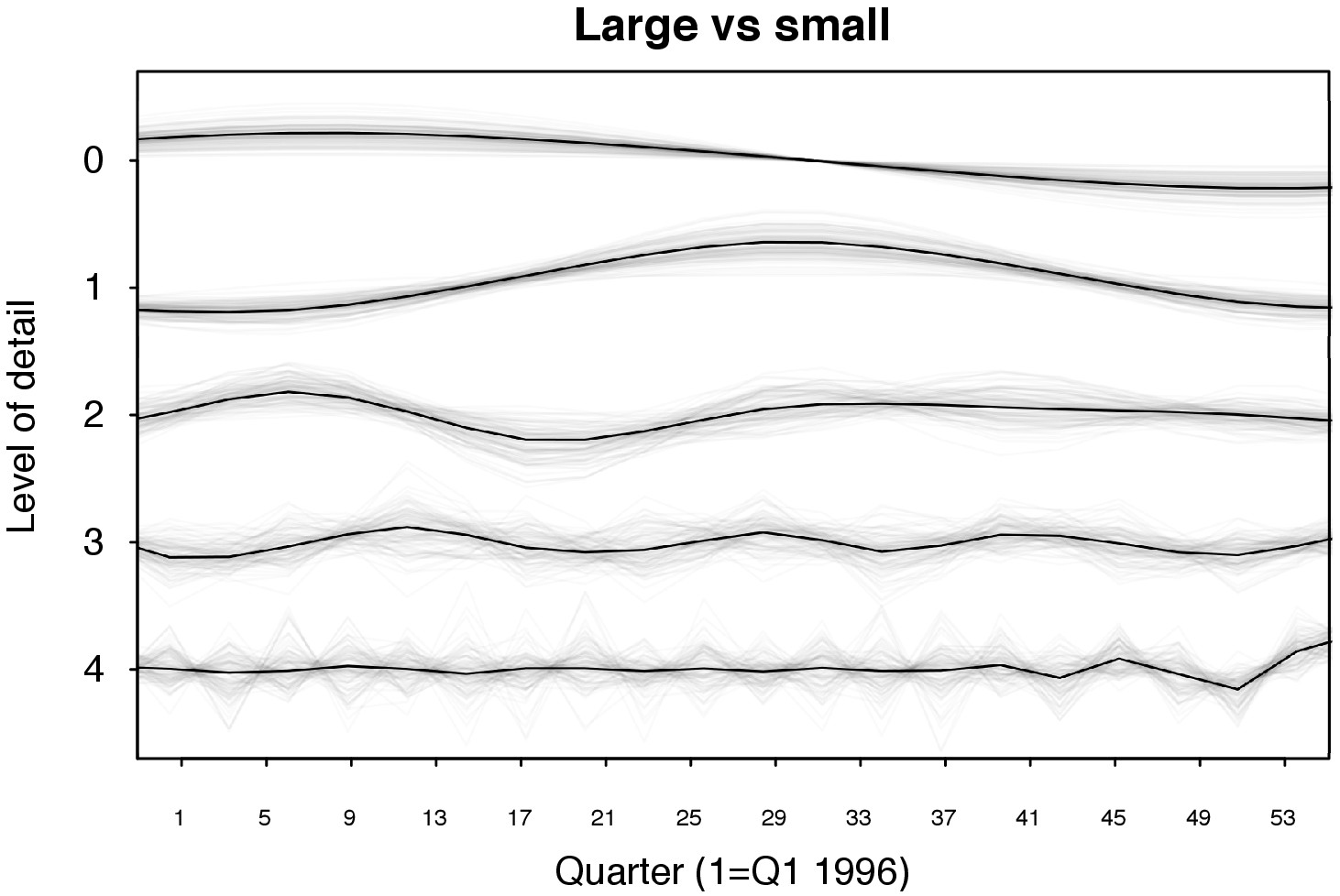} \\
\includegraphics[width=0.5\textwidth]{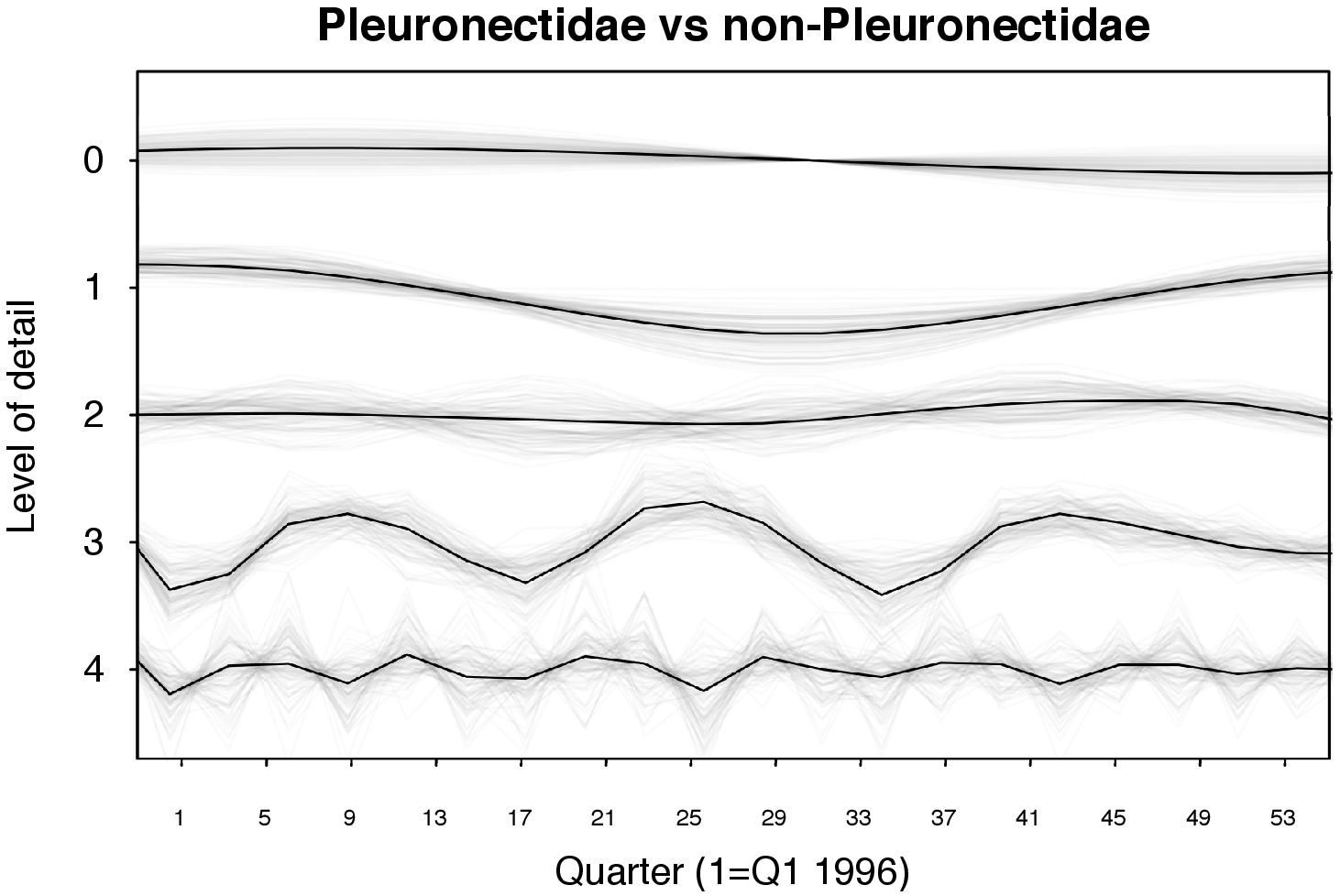} & \includegraphics[width=0.5\textwidth]{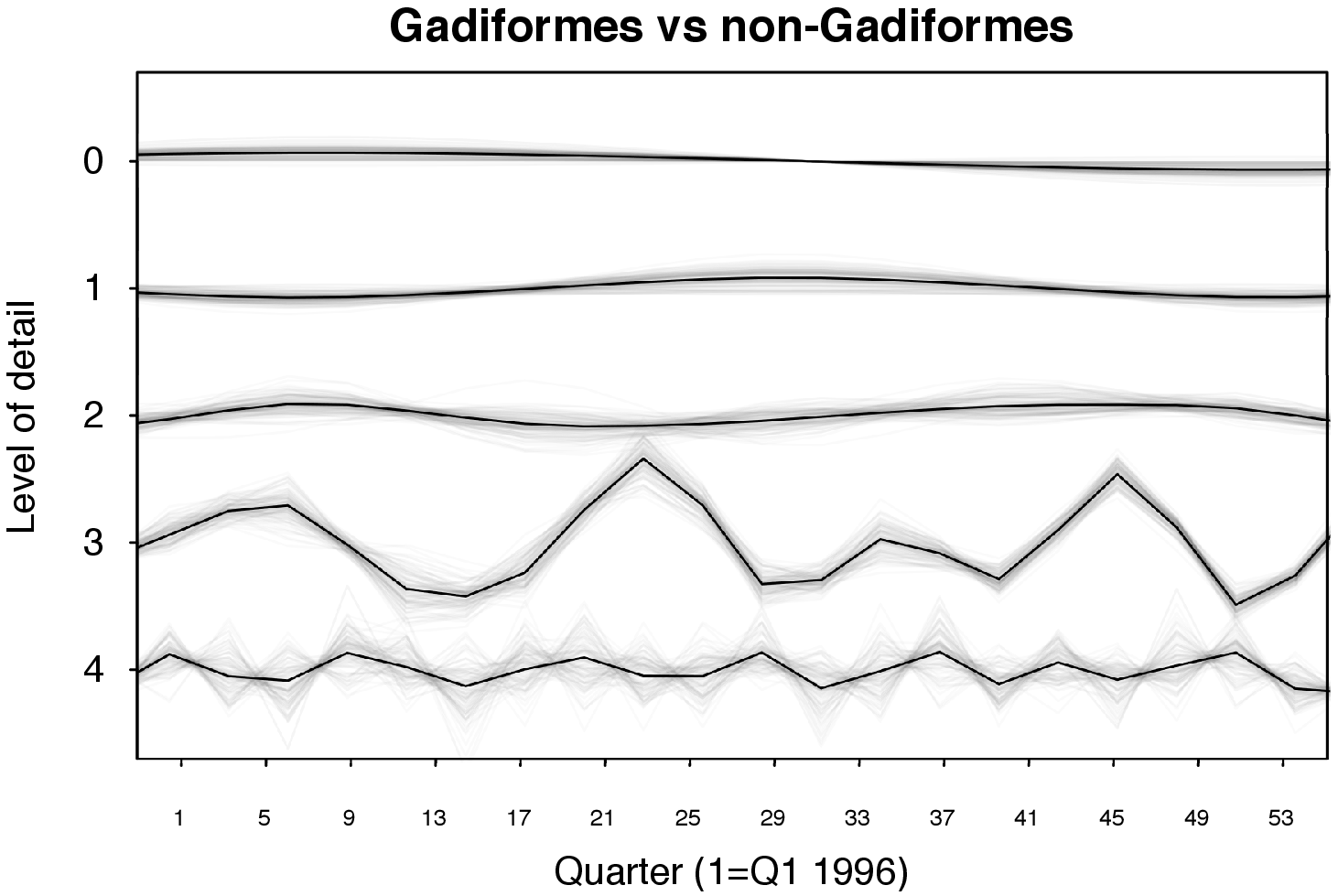} \\
\includegraphics[width=0.5\textwidth]{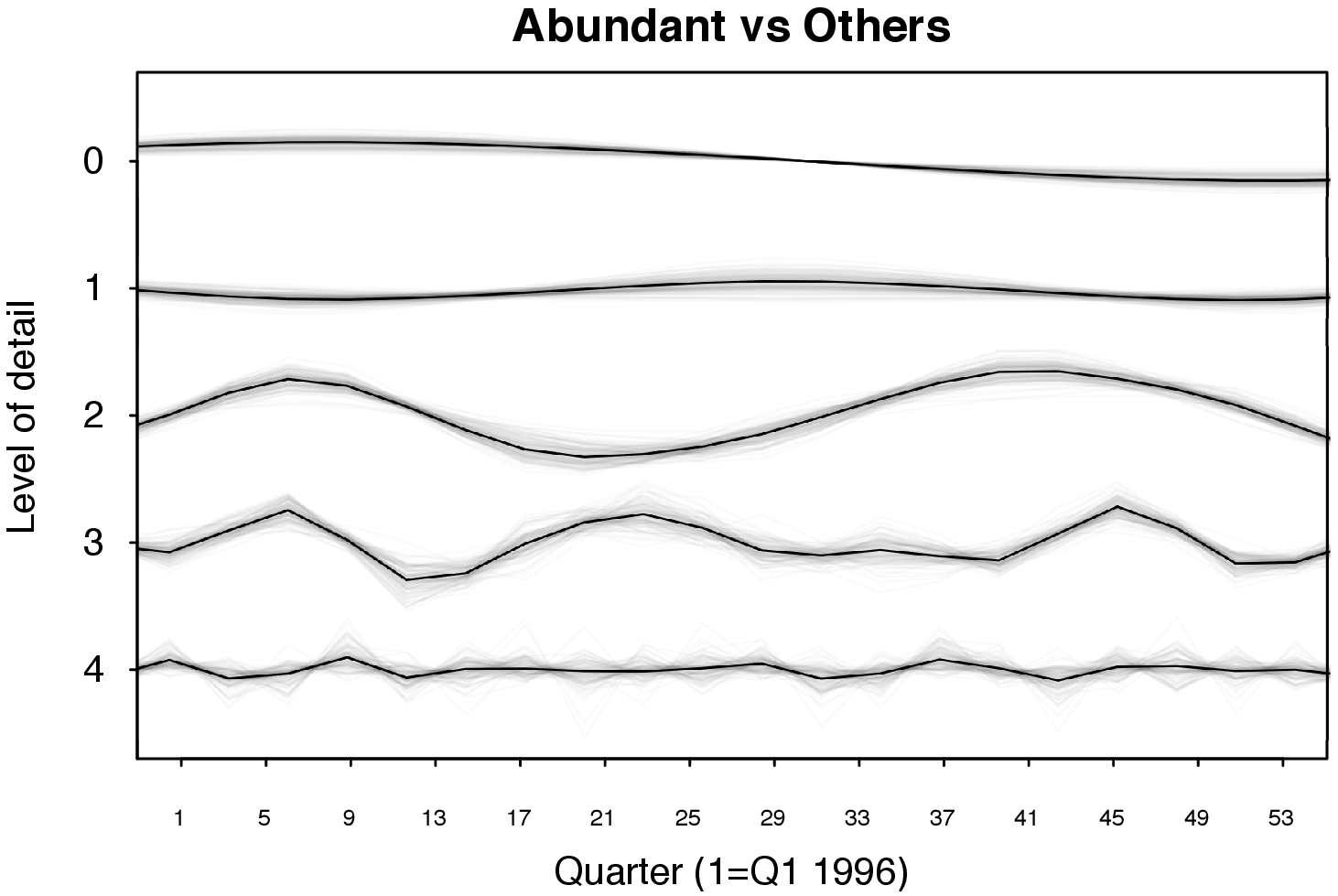} &  \\
\end{tabular}
\caption{Wavelet transforms for the 7 different models defined by the nesting structure of Figure \ref{nestingStructure}. The solid line shows the posterior median whilst the faint lines show 100 posterior sample wavelet transforms to give an idea as to the uncertainty.}
\label{all_WT}
\end{figure}

Taking the ZaNI-binomial model as the chosen model, Figure \ref{all_fits} shows fitted values for all models. Figure \ref{all_WT} shows the associated wavelet transforms. We can identify some seasonal behaviour at detail level 2 in Whiting vs Haddock, detail level 4 in Dab vs Plaice (after quarter 37), and detail level 3 in \textit{Pleuronectidae} vs non-\textit{Pleuronectidae} and \textit{Gadiformes} vs non-\textit{Gadiformes}. Detail level 4 corresponds most closely with yearly periodicity whilst detail level 3 corresponds to bi-yearly seasonality.\\

\begin{figure}[!p]
\begin{tabular}{cc} 
\includegraphics[width=0.5\textwidth]{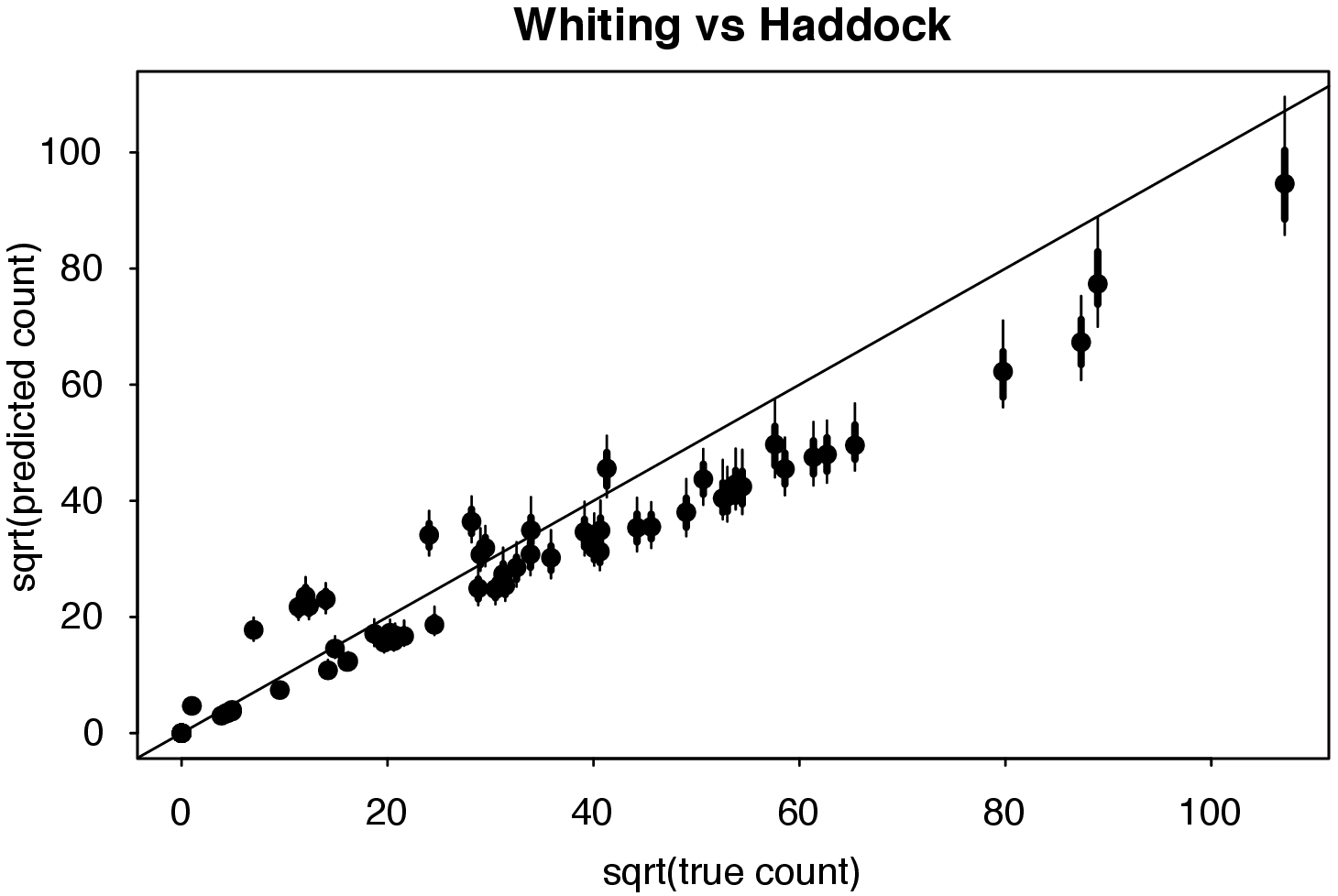} & \includegraphics[width=0.5\textwidth]{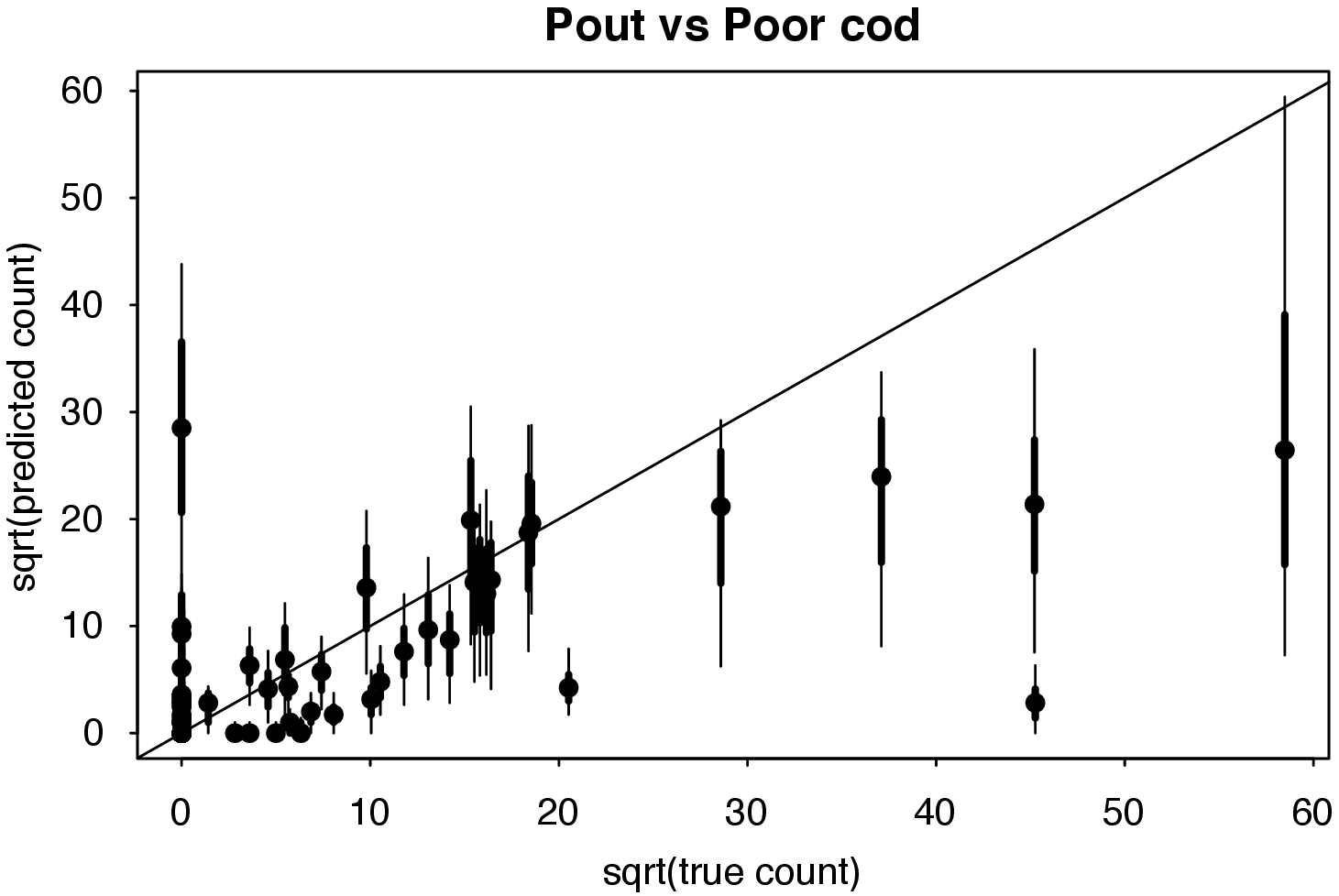} \\
\includegraphics[width=0.5\textwidth]{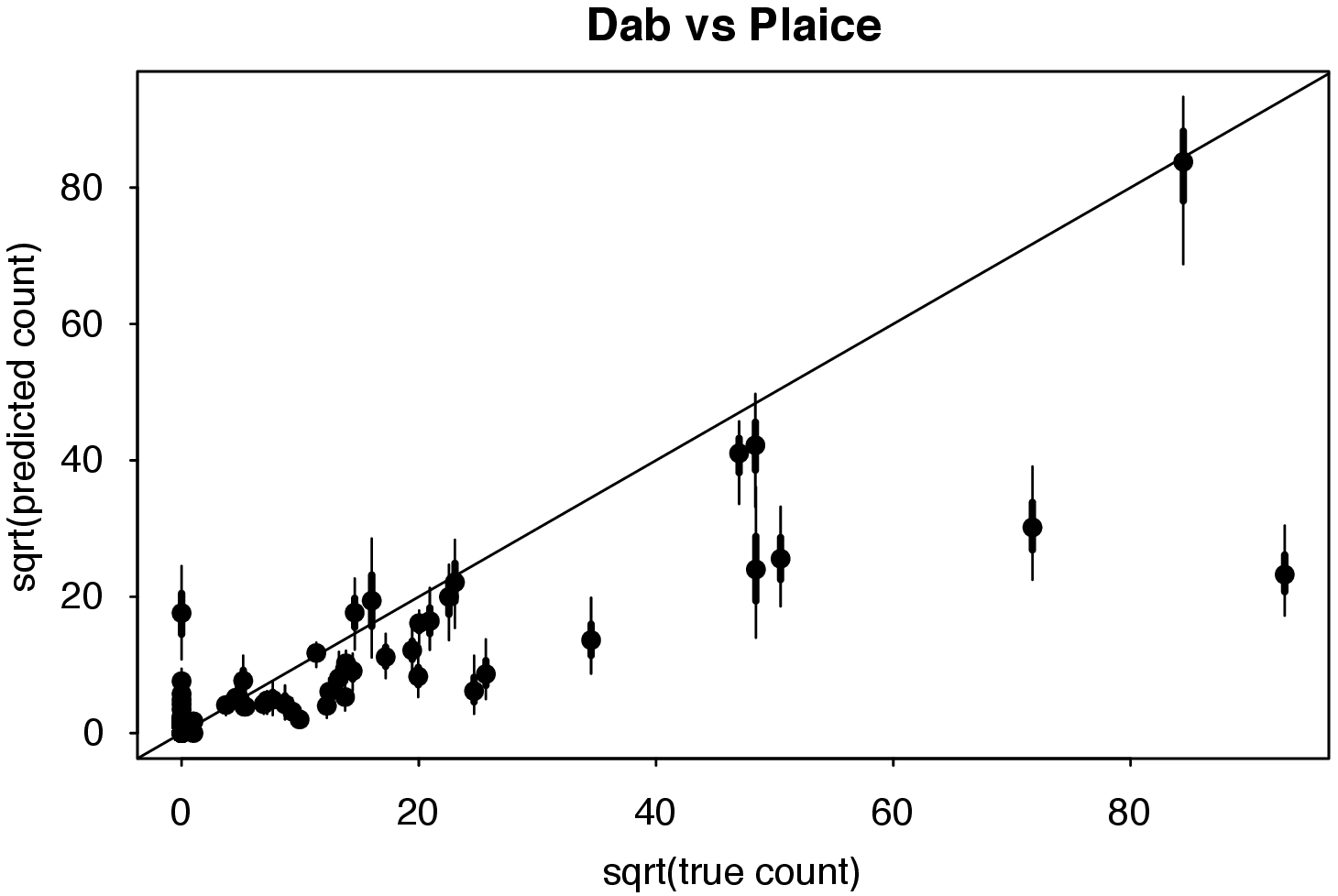} & \includegraphics[width=0.5\textwidth]{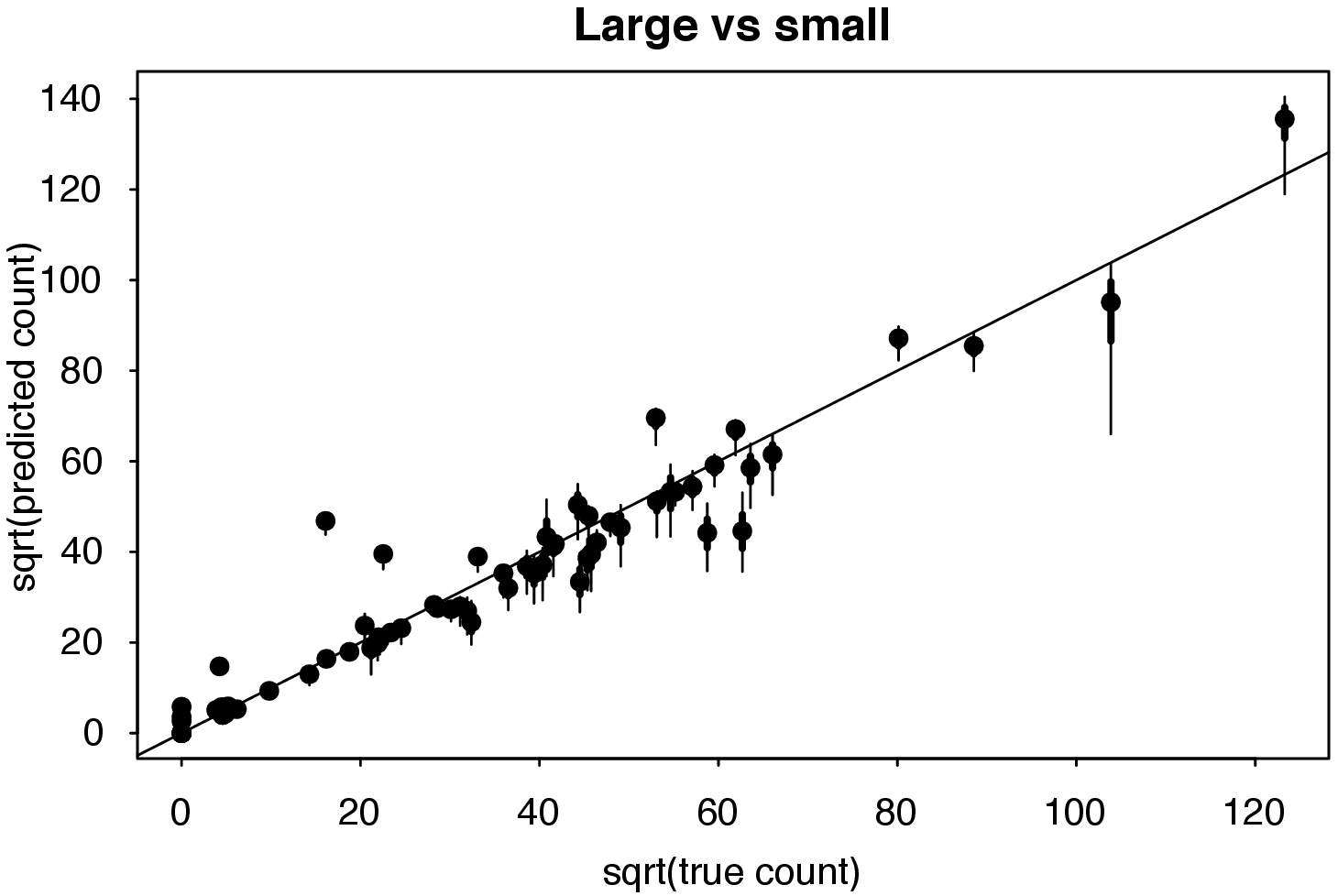} \\
\includegraphics[width=0.5\textwidth]{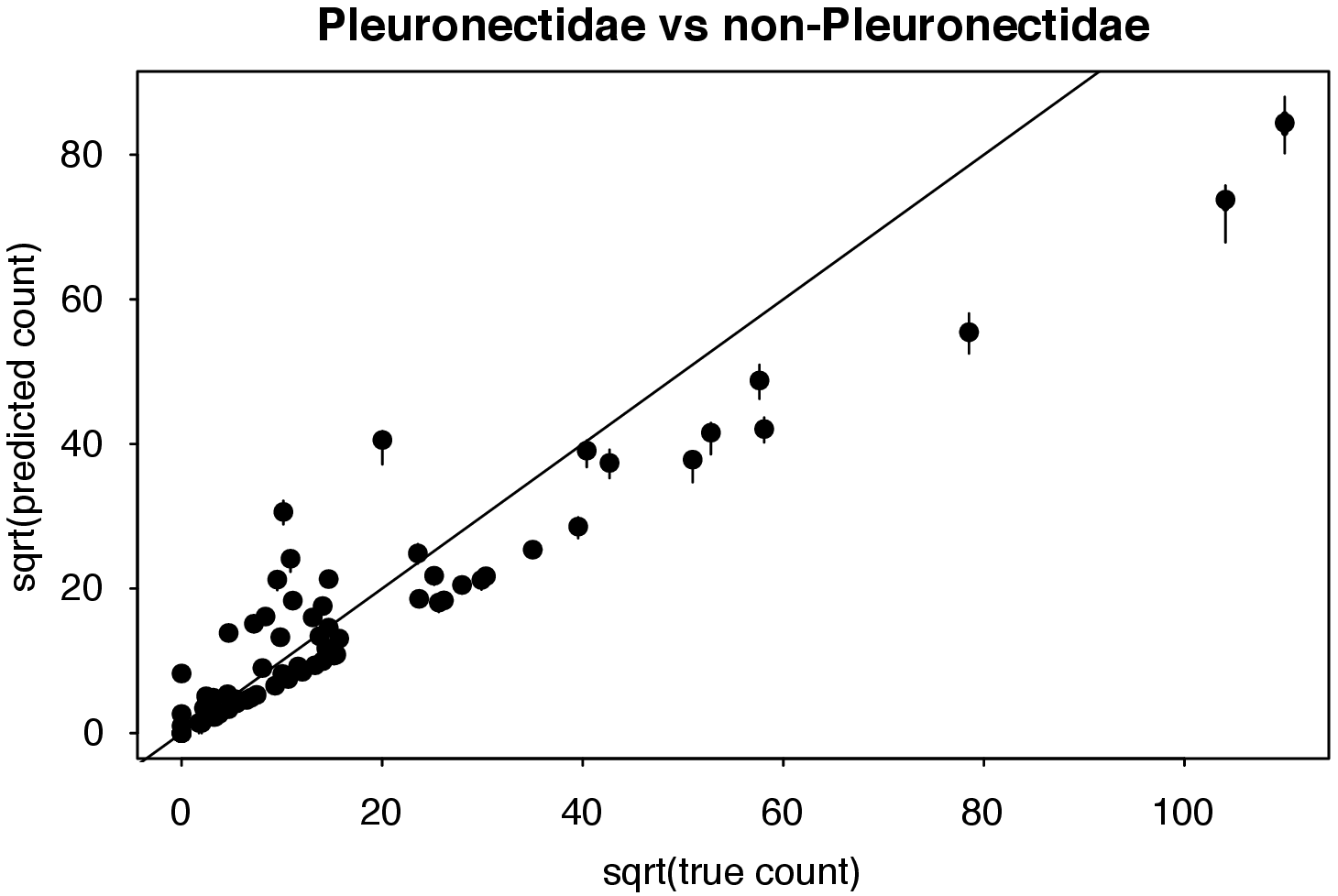} & \includegraphics[width=0.5\textwidth]{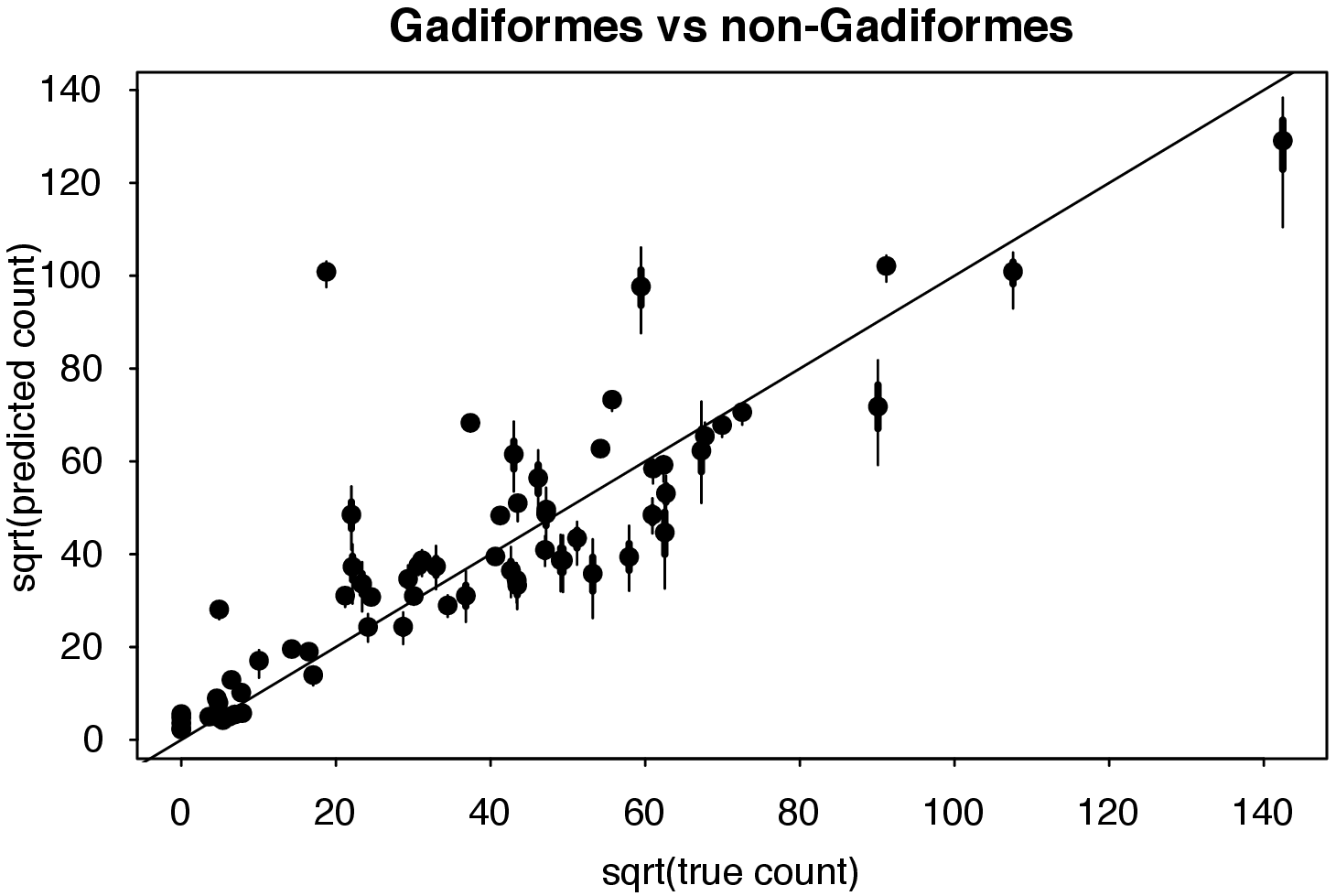} \\
\includegraphics[width=0.5\textwidth]{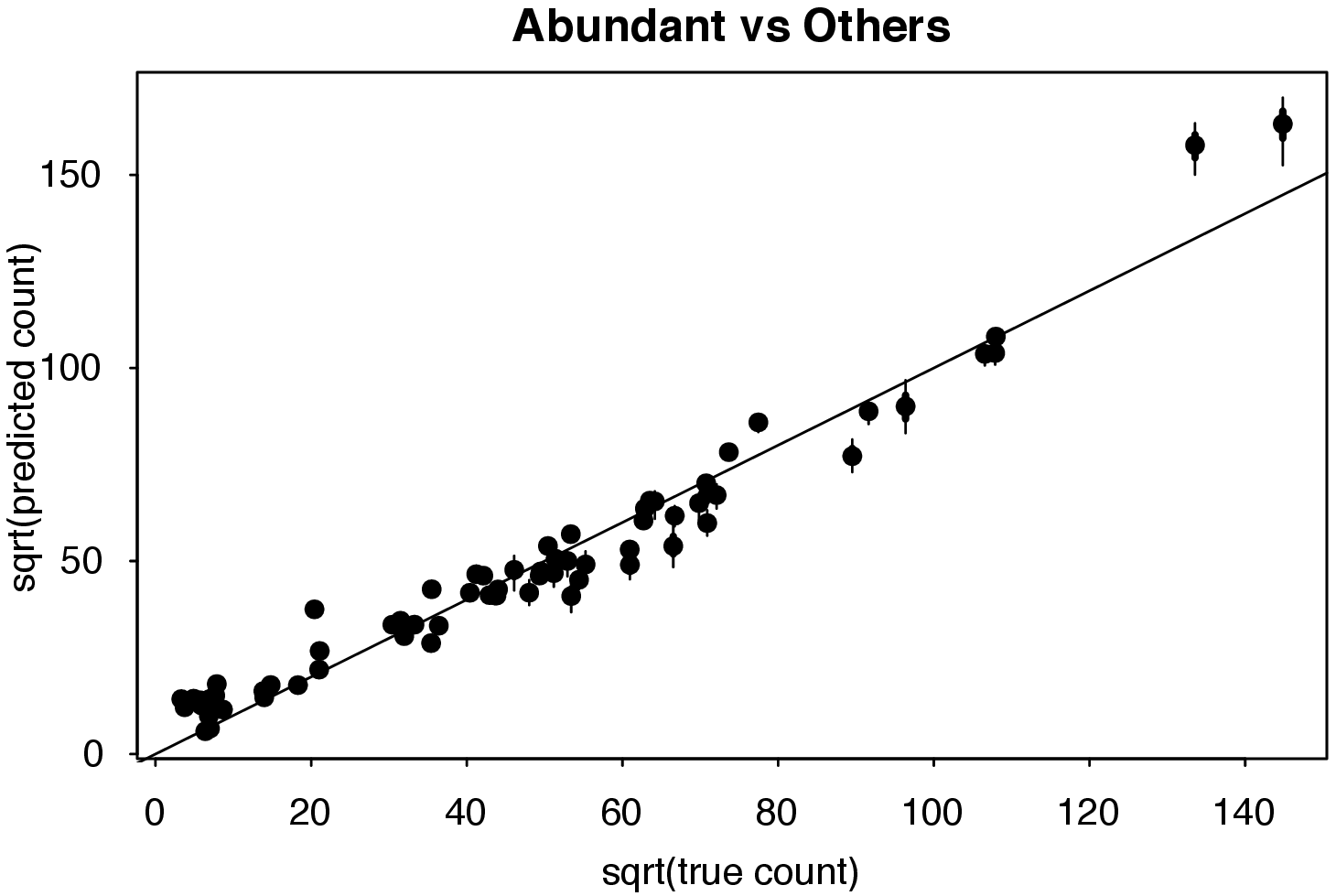} &  \\
\end{tabular}
\caption{Plots of predicted predictive values of $y$ (with 95\% CIs) versus true $y$ for a 10\% holdout data set. We plots these as square roots to avoid larger values dominating the plot.}
\label{all_holdout_preds}
\end{figure}

As a final test, we use a 10\% holdout data set to confirm that the ZaNI is a reasonable fit for out-of-sample data. For this test we do not use the random effects but predict from the values of $\mu_t$. Figure \ref{all_holdout_preds} shows the fitted vs posterior predicted $y$ values for each of the nesting models. The fit seems very good in all models but that of Pout vs Poor Cod which is notably the model for which there is the least evidence for the wavelet ZaNI-binomial (Table \ref{WAIC_table}). 

\section{Discussion}\label{discussion}

In this paper we have proposed and fitted a novel nested wavelet zero and $N$-inflated binomial model. The wavelet structure allows us to combine smoothing and frequency identification by using the wavelet transform. We have also used a novel shrinkage prior that has not been previously used in the wavelet literature. The nesting structure provides informative prior information on the important components of the data, and bypasses the slower and less informative traditional multinomial logit model. The performance of this model on these data seems acceptable, with the wavelet ZaNI-binomial being chosen most often via information criteria, and the out-of-sample fits showing good agreement with the left out data values.\\

At higher levels of the nesting structure (i.e. taxonomic order), we identified bi-yearly cycles of \textit{Gadiformes} and \textit{Pleuronectidae} which are likely to be a reflection of shifting fishing strategies and/or biological drivers such as spawning aggregations. For example, although most species spawn every year, certain year classes have a stronger recruitment and survival. Given that the majority of commercial species discards in the Irish Sea are from fish below the legal minimum landing size \cite{Borges2005}, the patterns observed could therefore be indicative of such phenomena in that a strong year class of \textit{Gadiformes} occurs every other year. \\

Annual fishing regulations such as quotas and seasonal closures (i.e. cod ÔboxÕ implemented in the Irish Sea around 2000 \cite{Kelly2006}), together with fish biology such as annual recruitment, could explain the annual cycles identified at the lower levels of the nested structure, for dab and plaice. \cite{Viana2012} identified both seasonal (year) cycles and long-term cycles for whiting and haddock discards in the Irish Sea. Although here we do not detect the annual cycles for these species, our estimate of 4 year cycles between haddock and whiting is consistent with \cite{Viana2012} that found a mean long-term cycle of 18 quarters for haddock and 25 for whiting, possibly representing recruitment pulses of these species. \\

Various improvements to the model seem appropriate. First, the wavelet basis functions do not provide exact estimates of frequency. This is the price we pay for having such flexible basis functions which can be fitted to data so easily. It may be that a different choice of basis functions may provide for better frequency identification. Second, the nesting structure we have used, whilst informed by the species relationships of the fish, may instead be learnt from the data. This seems somewhat similar to the process of learning a graphical model structure \citep{Yuan2007}. It may also be that other nesting structures of which we are not aware may fit the data better. Finally, the nesting structure does not provide for frequency identification of all individual species. It is quite simple to create estimates of the posterior proportion $p_k$ from simple transformations of the $\tilde{p}_k$, but this does not allow for wavelet transforms on all the individual species.\\

\section*{Acknowledgements}

The authors would like to thank James Sweeney and John Haslett for helpful discussions on the ZaNI-binomial distribution, and Claire Gormley for drawing our attention to the wavelet shrinkage prior. This work was partly supported by the Marine Research Sub-Programme of the Irish National Development Plan 2007 -- 2013 (grant PhD/FS/08/001).

5\bibliography{/Volumes/MacintoshHD2/BTSync/bibtex/library}
%\bibliography{/home/aparnell/BTSync/bibtex/library.bib}
%\bibliography{/Users/andrewparnell/BTSync/bibtex/library}
%\bibliography{library.bib}
\bibliographystyle{acmtrans-ims}

\end{document}